\documentclass[reprint,twocolumn,ap,prb,nofootinbib,showpacs]{revtex4-1}
\usepackage{graphicx}
\usepackage{hyperref}
\usepackage{amsfonts}
\usepackage{amsmath,amssymb}
\usepackage{bm}
\usepackage{color}

\begin{document}

\title{Soft superconducting gap in semiconductor-based Majorana nanowires}
\author{Tudor D. Stanescu}
\affiliation{Department of Physics and Astronomy, West Virginia University, Morgantown, West Virginia 26506, USA}
\author{Roman M. Lutchyn}
\affiliation{Station Q, Microsoft Research, Santa Barbara, California 93106-6105, USA}
\author{S. Das Sarma}
\affiliation{Condensed Matter Theory Center and Joint Quantum Institute, Department of Physics, University of Maryland, College Park, Maryland 20742, USA}

\begin{abstract}

We develop a theory for the proximity effect in superconductor--semiconductor--normal-metal tunneling structures, which have recently been extensively studied  experimentally,\cite{Mourik2012, Das2012, Deng2012, Fink2012, Churchill2013} leading to the observation of transport signatures consistent with the predicted zero-energy Majorana bound states.
We show that our model for the semiconductor nanowire having multiple occupied subbands with different transmission probabilities through the barrier reproduces the observed ``soft-gap'' behavior associated with substantial subgap tunneling conductance. We study the manifestations of the soft gap phenomenon both in the tunneling conductance and in local density of states measurements and discuss the correlations between these two quantities.
We emphasize that the proximity effect associated with the hybridization between low-lying states in the multiband semiconductor and the normal metal states in the lead is an intrinsic effect leading to the soft gap problem. In addition to the intrinsic contribution, there may be extrinsic effects, such as, for example, interface disorder, exacerbating the soft gap problem. Our work establishes the generic possibility of an ubiquitous presence of an intrinsic soft gap in the superconductor--semiconductor--normal-metal tunneling transport conductance induced by the inverse proximity effect of the normal metal.
\end{abstract}


\maketitle

\section{Introduction}
Interfaces between different materials are subjects of great current interest due to the possibility of exploiting various (emerging) properties of these structures. One such promising hybrid system is characterized by an interface between an ordinary $s$-wave superconductor (SC) and a semiconductor (SM) with strong spin-orbit coupling.\cite{Reich,Brouwer_Science, Wilczek2012} It has been shown recently\cite{Sau2010, Alicea2010, Lutchyn2010, Oreg2010} that, by combining these two conventional materials, one can realize in the SM a topological $p$-wave superconducting state that hosts exotic Majorana zero-energy modes.\cite{ReadGreen, Kitaev:2001, SrRu,FuKane} The defects carrying these modes obey non-Abelian braiding statistics,\cite{Moore1991, Nayak1996, ReadGreen, Ivanov} and can be utilized for topological quantum computation.\cite{TQCreview} The promise of engineering exotic physics using well-known generic materials has excited the experimental community, and there are significant experimental efforts aimed at finding Majorana zero-energy states in SM-SC hybrid structures,\cite{Mourik2012, Rokhinson2012, Das2012, Deng2012, Fink2012, Churchill2013} following the nanowire proposal of Refs.~\onlinecite{Lutchyn2010, Oreg2010}. Most of these experiments\cite{Mourik2012, Das2012, Deng2012, Fink2012, Churchill2013} pursued the detection of a zero-bias conductance peak associated with the Majorana modes through tunneling transport measurements and found the appearance of the peak at a finite magnetic field, as predicted.\cite{ZeroBiasAnomaly0,ZeroBiasAnomaly1,ZeroBiasAnomaly2,ZeroBiasAnomaly3,ZeroBiasAnomaly31, ZeroBiasAnomaly4,ZeroBiasAnomaly5,ZeroBiasAnomaly6, 1DwiresLutchyn2, ZeroBiasAnomaly61, ZeroBiasAnomaly7} However, a serious and persistent problem has been that, in addition to the peak, these experiments reveal a substantial subgap conductance, the origin of which has been mysterious, controversial, and highly problematic from the perspective of a theoretical interpretation of the experimental data. Furthermore, this soft-gap feature is the main roadblock for Majorana-based quantum computing proposals,\cite{AliceaBraiding, SauWireNetwork, ClarkeBraiding, TopologicalQuantumBus, BraidingWithoutTransport} which require a hard gap (i.e. vanishing subgap conductance) in the low-energy spectrum.\cite{ChengPRB'12}

In this article we develop a theory of the proximity effect in  superconductor--semiconductor--normal-metal (SC-SM-NM) nanostructures  that provides an explanation for the origin of the observed ``soft-gap,'' a feature originally designating the substantial subgap conductance revealed by the transport measurements on nanowire-SC hybrid structures.\cite{Mourik2012, Das2012, Deng2012, Fink2012, Churchill2013} We demonstrate that the very presence of a metallic lead coupled to the nanowire (i.e., even in the absence of a tunneling current) modifies the low-energy local density of states (LDOS) in the SC region of the wire and might generate a smooth subgap background, i.e., a soft gap. In other words, the normal metal itself produces an undesirable proximity effect on the SM wire, partially filling the proximity gap induced by the SC. In addition to transport, this effect could be revealed by, e.g., a scanning tunneling microscope (STM) measurement on a SC-SM-NM structure.
Using an effective model, we calculate the LDOS and the tunneling conductance in realistic settings and show that the states responsible for the soft gap are associated with low-lying occupied bands in the nanowire, which hybridize with the NM states and generate the subgap background due to the vastly varying transmission probabilities for different transverse subbands. Indeed, different transverse subbands are characterized by different hybridization couplings across the barrier with the highest (lowest)-occupied modes being most weakly (strongly) coupled to the NM lead. Given that the Majorana peak is due to states from the highest-occupied band, its experimental observation requires lowering the confining  potential barrier, which, in turn, increases the transmission probability for the lowest-occupied subbands. When this transmission probability becomes of order one (i.e. strong coupling limit for the given band), the hybridization with the metal generates the subgap background.
We note that previous studies\cite{AkhmerovPRL'11,PradaPRB'12,PientkaPRL'12,RainisPRB'13}
have concluded that one cannot exactly reproduce the substantial background subgap density of states seen in the experiments.\cite{Mourik2012, Das2012, Deng2012, Fink2012, Churchill2013} The origin of the soft-gap was attributed to some extrinsic mechanisms, i.e. interface disorder.\cite{TakeiPRL'13} In this article, we explicitly show the role of multiband occupancy and of band-selective SM-NM coupling in the emergence, even in a clean system, of the soft-gap and argue that this multiband  intrinsic model provides a natural explanation of the experimental results.

\begin{figure}[tb]
\begin{center}
\includegraphics[width=0.48\textwidth]{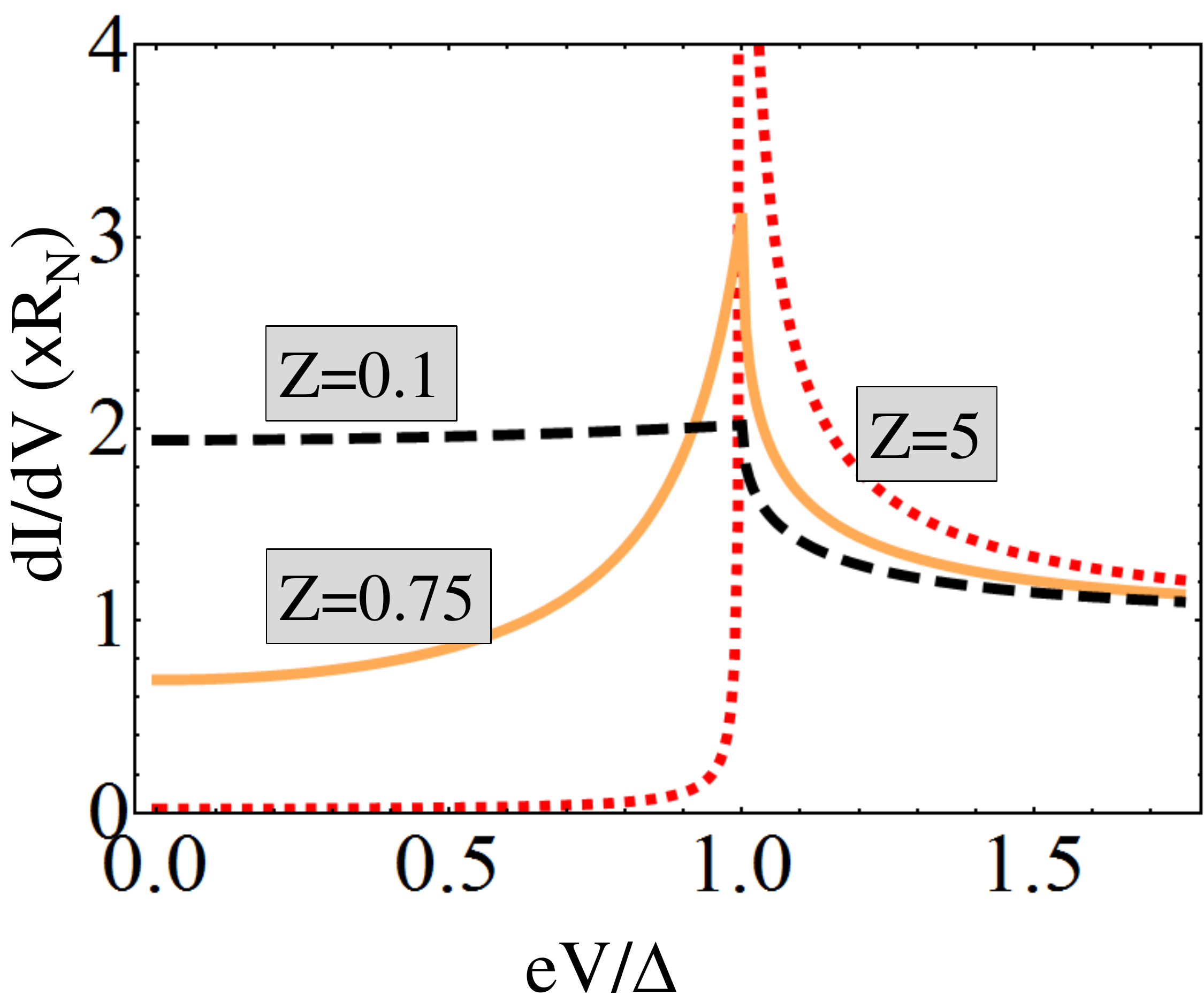}
\vspace{-8mm}
\end{center}
\caption{(Color online)  Differential conductance versus voltage for different values of the barrier strength $Z$. The curves, illustrating the main result of the BTK theory, correspond to zero temperature and are normalized by multiplying $dI/dV$ with the normal-state resistance $R_N$ (see Ref. \onlinecite{BTK}). In the strong-barrier limit ($Z\rightarrow\infty$) $dI/dV$ is proportional to the density of states in the superconductor, but the two quantities become completely unrelated in the weak-barrier regime.}
\vspace{-4mm}
\label{FFig1}
\end{figure}

\section{The intrinsic soft-gap mechanism: A qualitative picture}

In this section we provide a qualitative picture of the soft-gap mechanism that summarizes our technical results. For clarity, we would like to emphasize from the very beginning two key points. 1) This is an ``intrinsic'' mechanism that links the emergence of a smooth in-gap background -- the soft-gap -- to the presence of a metallic lead strongly coupled to the semiconductor nanowire. In addition, in real structures, there may be
``extrinsic'' contributions generated, for example,  by interface disorder and potential impurities.\cite{Stanescu2011,TakeiPRL'13} In an isolated SM-SC system these states generate sharp subgap spectral features that depend on the specific disorder realization. However, the coupling of the nanowire to a normal-metal lead will broaden these sharp features, leading to a smooth background. Thus, in general, both the intrinsic and the extrinsic mechanisms are expected to contribute to the soft-gap feature.
 2) The soft gap occurs due to the coupling of the nanowire to a normal-metal lead used to probe the active SM-SC system in a transport measurement. However, this is not an exclusive transport feature  (e.g., a property of the differential conductance $dI/dV$), but also characterizes the equilibrium properties of a NM-SM-SC system (e.g., the local density of states).  In other words, if $dI/dV$ is characterized by a soft-gap,  the LDOS (inside the barrier region) in the absence of any charge current will also be characterized by a soft-gap. The features associated with these different manifestations of the soft-gap are in one-to-one correspondence.

The soft-gap phenomenon, as manifested in transport measurements, can be understood within the Blonder-Tinkham-Klapwijk (BTK) theory,\cite{BTK} which describes the crossover from metallic to tunnel junction behavior. Let us consider a typical system currently used in Majorana experiments (see, for example, Ref.\onlinecite{Mourik2012}). A certain segment of a SM nanowire is proximity-coupled to an $s$-wave SC, while the remaining part of the wire is coupled to a NM lead through an Ohmic contact. The superconducting segment, which represents the ``active system,'' is separated from the normal part (which represents the probe) by a potential barrier $V(x)$. For simplicity, let us assume that $V(x)=w\delta(x)$. For a given SM nanowire band (say $n$), the transparency of the potential barrier depends on the strength of the potential, $w$, as well as the characteristic Fermi velocity, $v_F^{(n)}$. More specifically, in the normal state (i.e., above some critical temperature) the transmission coefficient is\cite{BTK} $T_n=1/(1+Z_n^2)$, where $Z_n=w/\hbar v_F^{(n)}$ is a band-dependent dimensionless barrier strength. As shown in Fig. \ref{FFig1}, the contribution of band $n$ to the differential conductance $dI/dV$ depends critically on the barrier strength $Z_n$. In the strong-barrier ($Z_n\rightarrow\infty$), i.e., low transparency ($T_n\ll 1$),  limit the differential conductance is proportional with the SC density of states, i.e., the density of states characterizing the {\em decoupled} SC segment separated from the probe by an infinite barrier. If there are no in-gap states (as assumed in the calculation shown in Fig. \ref{FFig1}), $dI/dV$ will be characterized in this limit by a ``hard'' gap. By contrast, for  low barriers ($Z_n<1$), i.e., high transparencies ($T_n \lesssim 1$), the differential conductance corresponding to values of the bias voltage inside the SC gap (i.e., $eV<\Delta$) becomes finite. As a result, upon lowering the barrier height,  the $dI/dV$ hard gap becomes  soft and, eventually, is completely filled in (see Fig. \ref{FFig1}).

Based on the above observations, one concludes that the soft-gap phenomenon has to be associated with a high transparency barrier. However, in order to obtain any sharp feature in the tunneling conductance (such as the expected Majorana zero-bias peak), the potential barrier should be in the low-transparency limit for the top occupied SM band, i.e., the band with the lowest $v_F^{(n)}$ value. This top occupied band is, in fact, responsible for the Majorana physics (hence it is sometimes called the Majorana band). Consequently, the soft-gap phenomenon observed in the recent Majorana experiments is based on a mechanism that has two critical ingredients: (i) multiband occupancy and (ii) band-selective coupling to the external probe. The height of the potential barrier is experimentally adjusted to ensure the visibility of the zero-bias peak (ZBP). This corresponds to the top (Majorana) band being in the weak-coupling (low-transparency) limit. However, the barrier potential cannot be too strong because this would lead to a ZBP with vanishingly small weight (i.e., it leads to the disappearance of the ZBP, as we will explicitly show below). Consequently, the potential barrier height has to be within a certain optimal window. These optimal potential values correspond to a highly transparent barrier for the low-energy occupied bands. This leads to a finite $dI/dV$ inside the SC gap, i.e., to the soft-gap phenomenon. Note that single-band models\cite{AkhmerovPRL'11,ScattMat2} do not contain the main ingredients identified above and, consequently, cannot reproduce the background subgap density of states associated with the soft-gap feature.

Next, we turn to the equilibrium manifestation of the soft-gap as revealed by the density of states.
As discussed above, transport measurements provide information on the density of states of the active system {\em only} in the weak tunneling limit ($Z_n \gg 1$, $T_n \ll 1$). However, in a multiband wire this condition might not be satisfied for all  occupied bands, since the transparency of the barrier is band dependent. To gain a better understanding of the mixed regime that characterizes systems with both high and low transmission bands, we also calculate the local density of states, a thermodynamic quantity that can be measured using, for example, STM. Since some of the bands are highly transparent, it is critical to take into account the effect of the normal lead in a non perturbative  way.
The coupling of a low density SM to an $s$-wave SC and a NM lead generates proximity effects that can be understood within the Green's function framework. Due to the SM-SC proximity effect, the SM wire acquires a spectral gap. This SC proximity effect can be taken into account through a boundary self-energy $\Sigma_{ij}^{\rm SC}(\omega)$ that is due to the exchange of particles between the two subsystems. In addition, the coupling to the normal metal leads to a SM-NM proximity effect that accounts for the hybridization of SM and NM states and can be described by an interface self-energy $\Sigma_{ij}^{\rm NM}(\omega)$ containing both real and imaginary contributions. Thus the density of states in the SM is determined by a combined effect of the SC and the NM. The effect of the normal lead is particularly important for the density of states below the induced SC gap, where nominally (i.e., in an active system {\em without} any lead) the quasiparticle density of states would be zero. In the presence of the SM-NM coupling, the in-gap density of states acquires a finite correction induced by the NM and, consequently, the quasiparticle gap becomes soft.

\begin{figure}[t]
\begin{center}
\includegraphics[width=0.48\textwidth]{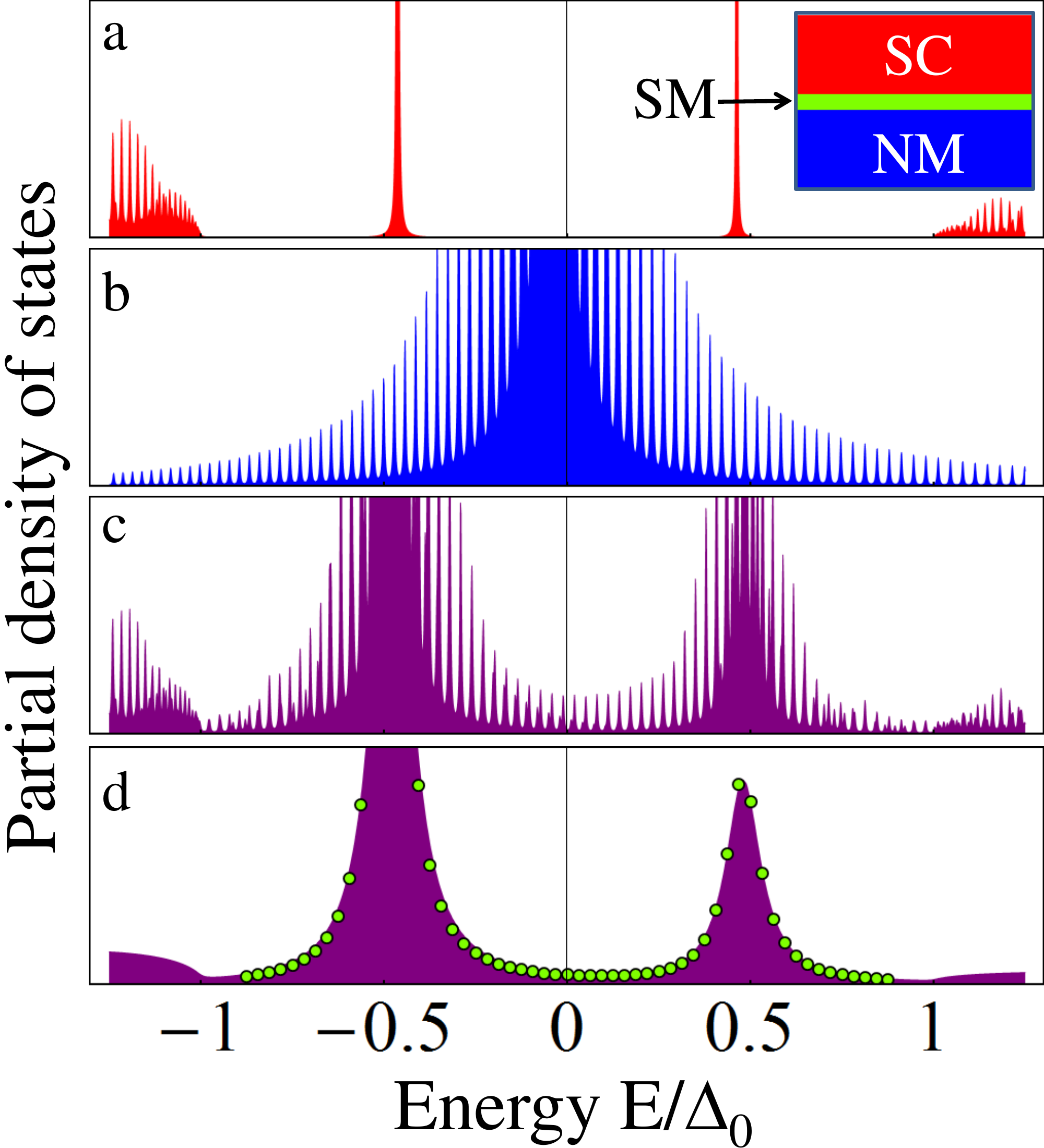}
\vspace{-7mm}
\end{center}
\caption{(Color online) Contribution of a hybridized  SM state with bare energy $E_0=0$ to the LDOS integrated over the semiconductor nanoribbon volume in a SC-SM-NM structure (lateral view shown in the inset). (a) SM-SC coupling ($\gamma_{SC}\neq 0$, $\gamma_{NM}=0$). The sharp peaks correspond to Bogoliubov quasiparticles with energies given by the induced SC gap $\Delta_{ind}\approx 0.5\Delta_0$. (b) SM-NM coupling ($\gamma_{SC} = 0$, $\gamma_{NM}\neq0$). Each peak corresponds to a SM-NM hybrid state having most of its weight  inside the NM and a small tail inside the SM ribbon. (c) Proximity-coupled NM-SM-SC structure ($\gamma_{SC}\neq 0$, $\gamma_{NM}\neq 0$).  The hybrid states with energies within the induced SC gap are responsible for the soft-gap phenomenon. The weight of the corresponding peaks depends on the SM-NM coupling strength  $\gamma_{NM}$. (d) Same as in panel (c) but for a system with large SC and NM components (with quasi-continuum spectra). The green circles are given by an effective theory obtained by integrating out the SC and NM degrees of freedom (see Sec. \ref{theory}).}
\vspace{-0mm}
\label{FFig2}
\end{figure}

To gain further insight, let us consider a hybrid system consisting of a semiconductor nanoribbon sandwiched between a superconductor and a normal metal, as shown in the inset of Fig. \ref{FFig2}(a). We focus on the contribution to the local density of states (LDOS) integrated over the semiconductor volume of a single SM state (with energy $E_0=0$) that hybridizes with NM and SC states. The effective SM-SC and SM-NM couplings are $\gamma_{SC}$ and $\gamma_{NM}$, respectively (see Sec. \ref{theory} for the exact definition of these quantities), and the bulk SC has a gap $\Delta_0$. The results are shown in Fig. \ref{FFig2}.
Panel (a) corresponds to a SM ribbon proximity coupled to a mesoscopic SC only, i.e., $\gamma_{SC}\neq 0$ and $\gamma_{NM}=0$. The only contributions within the bulk SC gap ($\Delta_0$), i.e., the two sharp peaks, arise from two Bogoliubov quasiparticles with energies $\pm\Delta_{ind}$ corresponding to the proximity-induced SC gap. Note that hybridization with the (discrete) SC states pushes some of the spectral weight above $\Delta_0$.
Panel (b) corresponds to a SM nanoribbon coupled to a mesoscopic NM ($\gamma_{SC}= 0$, $\gamma_{NM}\neq 0$).   Each peak corresponds to a hybrid state that has most of its spectral weight inside the NM and a small tail (of weight proportional to the peak height) inside the SM. Panel (c) represents a proximity-coupled NM-SM-SC structure ($\gamma_{SC}\neq 0$, $\gamma_{NM}\neq 0$).  Individual peaks correspond to low-energy states characterized by nonzero spectral weight inside all three components of the system. The soft induced gap emerges as an energy-dependent modulation of the spectral weight in the SM tails with maxima near  $\pm\Delta_{ind}$. The heterostructure corresponding to panel (d) is the same as in (c), but has large NM and SC components with  quasi continuum spectra. The green circles are determined using an effective theory obtained by integrating out the SC and NM degrees of freedom (see Sec. \ref{theory} for details).

We conclude that the soft-gap is due to the emergence of hybrid states with energies within the induced SC gap and finite spectral weight inside the SM nanoribbon. The connection with the transport picture presented above is straightforward. Indeed, for strong-enough SM-NM coupling (which corresponds to Ohmic contact between the SM wire and the NM lead in an experimental hybrid structure), hybrid states with subgap energies extend throughout the whole system and participate to transport. This results in a finite subgap $dI/dV$, i.e., the soft-gap behavior.

We end this section with two important observations. First, we emphasize that in the experimentally relevant configuration the coupling between the SM states and the metallic lead is controlled by two different parameters: the coupling at the SM-NM interface ($\gamma_{NM}$) and the strength/transparency of the potential barrier (i.e., $Z_n$ or $T_n$). The SM-NM coupling has to be strong enough to ensure Ohmic contact between the  SM nanowire and the metallic lead. Otherwise, the interface will correspond to a tunnel junction and the measured $dI/dV$ will be proportional to the LDOS on the SM side of the SM-NM junction, which is different from the LDOS at the end of the superconducting segment of the wire (the active system), i.e., the quantity that carries information about Majorana physics, including the ZBP. In other words, if the coupling across the SM-NM interface is weak, the Majorana band, consisting of states that are confined within the SC segment of the wire, will be completely decoupled  from the lead and will not generate any measurable signature. The other parameter -- the strength of the potential barrier -- is responsible for the band-selective coupling between the SM states and the probe. In a multiband system, the optimal condition for the observation of the Majorana ZBP may correspond to a high-transparency barrier for the low-energy occupied bands, which leads to the soft-gap phenomenon. Note that the relative transparencies of  these bands depend strongly on the profile of the barrier potential. In particular, for experimentally relevant wide barriers (with characteristic widths of the order of $100$ nm)  the high- (low-) transparency condition for the low-energy (Majorana) bands is naturally satisfied, while for a narrow square potential barrier the high-transparency condition $T_n\lesssim1$ may not be satisfied in a system with a small number of occupied bands. Consequently, multiband calculations based on a narrow square potential (e.g., Ref. \onlinecite{PientkaPRL'12}) may not capture the soft-gap phenomenology.

Our second comment concerns the role of extrinsic effects, in particular that of interface disorder.\cite{TakeiPRL'13}  The presence of disorder leads to the proliferation of low-energy subgap states. If the disorder is characterized by a short characteristic length scale, it strongly affects the low-energy bands with $1/k_F^{(n)}$ comparable with this characteristic length.  Consequently, the (isolated) active system will have a nonvanishing subgap density of states characterized by discrete spectral features. Upon turning on the coupling to the probe, these features evolve into a smooth background. We note that, at finite temperature, thermal effects will contribute to an additional broadening of the spectral features (e.g., of the order $k_B T$ for the differential conductance).
In general, both the extrinsic effect and the intrinsic mechanism described in this work are expected to contribute to the soft-gap phenomenon with relative contributions that depend on the details of the system (e.g., quality of the wire and of the SM-SC interface, barrier potential profile,  barrier height, etc.). As demonstrated by this study, eliminating the extrinsic contributions (i.e., engineering clean structures) does not guarantee the absence of the soft-gap feature. Also, while increasing the gate potential can result in a hard gap, this also suppresses the weight of the Majorana-induced  ZBP, which may become unobservable. As we will discuss in more detail in Sec. \ref{results}, the practical solution for obtaining a hard gap {\em and} an observable ZBP involves  engineering clean structures plus low occupancy and narrow potential barriers.

\section{Theoretical modeling} \label{theory}

\subsection{Effective tight-binding model}

The Hamiltonian describing the NM-SM-SC heterostructure has the generic form\cite{Stanescu2011}
\begin{equation}
H^{\rm tot} = H^{\rm SM} + H^{\rm V} + H^{\rm SC} + H^{\rm NM} + H^{\rm SM-SC} + H^{\rm SM-NM},                  \label{Htot}
\end{equation}
where the terms on the right-hand side represent the contributions from the SM wire, gate potential, bulk SC, NM contact, SM-SC coupling, and SM-NM coupling, respectively. These contributions are incorporated using realistic tight-binding models that include on-site and nearest-neighbor terms defined on cubic lattices with a lattice constant $a$. The SM wire has a rectangular cross section and characteristic length scales $L_x \gg L_y \gg L_z$. The thin-wire assumption ($L_z\ll L_y$)  is made to avoid complications arising from the  thickness dependence of the SC proximity effect which has already been studied elsewhere.\cite{Stanescu2013a} The relevant parameters are the lattice constant  $a = 6.48$ \AA, the lengths of the superconducting ($L_S$) and normal ($L_N$) segments of the wire ($L_S+L_N = L_x$),  the height of the potential barrier $V_b^{max}$,  the nearest-neighbor hopping
$t_{SM}=5.67$ eV, the Rashba spin-orbit coupling $\alpha_R=200.0$ meV$ $\AA, and the Zeeman splitting $\Gamma$.
The semiconductor wire is described by a tight-binding model with nearest-neighbor hopping on a cubic lattice and includes the effects of spin-orbit coupling. Explicitly,
\begin{align}
H^{\rm SM} = H_0 +H_{\rm SOI} = -t_{SM}\sum_{{\bm i}, {\bm \delta}, \sigma}c_{{\bm i}\!+\!{\bm \delta}\sigma}^{\dagger}c_{{\bm i}\sigma} -\mu \sum_{{\bm i}, \sigma} c_{{\bm i}\sigma}^{\dagger}c_{{\bm i}\sigma}  \nonumber \\
+ \frac{i \alpha}{2}\sum_{{\bm i},{\bm \delta}}\left[ c_{{\bm i}+{\bm \delta}_x}^{\dagger}{\sigma}_y c_{{\bm i}} -  c_{{\bm i}+{\bm \delta}_y}^{\dagger}{\sigma}_x c_{{\bm i}} + {\rm h.c.} \right],  \label{Eq2}
\end{align}
 where $H_0$ includes the first two terms and describes nearest--neighbor hopping on a simple cubic lattice with amplitude $t_{SM}$ for a system with chemical potential $\mu$ and the last term represents the Rashba spin-orbit interaction. Here, ${\bm i}=(i_x, i_y,i_z)$ labels the lattice sites and ${\bm \delta}\in\{{\bm \delta}_x, {\bm \delta}_y, {\bm \delta}_z\}$ are nearest--neighbor position vectors. In the calculations we consider a cubic lattice with a lattice constant $a = 6.48$ \AA and we use a value for the hopping matrix element $t_{SM}=5.67$ eV, which corresponds to an effective mass $m_{eff}=0.016m_0$ that characterizes the conduction band of InSb. The coefficient of the Rashba spin-orbit coupling is $\alpha_R=\alpha a=200.0$ meV \AA. The wires considered in the numerical calculations have a rectangular profile with $(L_x, L_y, L_z)$ lattice sites along the $x$, $y$, and $z$ direction, respectively. The wire of length $L_x = L_N+L_{S}=3859$ lattice sites has a normal segment of length 1.1 $\mu$m ($L_N=1700$)  partly covered by the normal metal and a segment of length $a L_S=$ 1.4 $\mu$m coupled to the SC  [see Fig. \ref{FFig3}(a)]. The width of the wire is 110 \AA   ($L_y=170$) and the thickness is 25 \AA ($L_z=38$).  In the presence of a magnetic field oriented parallel to the wire,  the Hamiltonian (\ref{Eq2}) is supplemented by a Zeeman term $H_{\Gamma}= \Gamma \sum_{{\bm i}}[c_{{\bm i}\uparrow}^{\dagger}c_{{\bm i}\downarrow} + h.c.]$, where $\Gamma$ is the Zeeman splitting.
 The confining gate potential is described by a purely local contribution to the Hamiltonian,
\begin{equation}
H^{\rm V} = \sum_{{\bm i}, \sigma} V_b(i) c_{{\bm i}\sigma}^{\dagger}c_{{\bm i}\sigma},          \label{Eq3}
\end{equation}
with $V_b(x) = V_b^{max}\exp[{-(x-x_b)^2/w_b^2}]$, where $V_b^{max}$ is the height of the gate-induced potential barrier, $x_b$ gives the location of barrier [e.g., $x_b=$0.97 $\mu$m in Fig. \ref{FFig3}(a)], and  $w_b=52$ nm is the characteristic width of the barrier.

The superconductor is described at the mean-field level using a simple tight-binding Hamiltonian characterized by a constant pairing amplitude $\Delta_0=1.5$ meV.\cite{Stanescu2013b} Explicitly, we have
\begin{equation}
H^{SC} = \sum_{{\bm i}, {\bm j}, \sigma} \left(t_{ij}^{sc} - \mu_{SC}\delta_{\bm i \bm j}\right) a_{{\bm i}\sigma}^\dagger a_{{\bm j}\sigma} + \Delta_0\sum_{\bm i} (a_{{\bm i}\uparrow}^\dagger a_{{\bm i}\downarrow}^\dagger + a_{{\bm i}\downarrow}a_{{\bm i}\uparrow}),  \label{Eq4}
\end{equation}
where $i$ and $j$ label SC lattice sites, $a_{i\sigma}^\dagger$ is the creation operator corresponding to a single-particle state with spin $\sigma$ localized near site $i$, the hopping matrix elements are nonzero, $t_{ij}^{sc}=t_{SC}$, only if $i$ and $j$ are nearest neighbors, and $\mu_{SC}$ represents  the chemical potential of the SC. Similarly, the normal-metal lead is modeled using a tight-binding Hamiltonian with nearest-neighbor hopping on a cubic lattice with the same lattice constant $a = 6.48$ \AA. Explicitly,
\begin{equation}
H^{NM} = \sum_{{\bm i}, {\bm j}, \sigma} \left(t_{ij}^{nm} - \mu_{NM}\delta_{\bm i \bm j}\right) b_{{\bm i}\sigma}^\dagger b_{{\bm j}\sigma}, \ \label{Eq5}
\end{equation}
where $t_{ij}^{nm}=t_{NM}$ if $i$ and $j$ are nearest neighbors and zero otherwise.

The coupling terms in Eq.~(\ref{Htot}) include nearest-neighbor hopping across the SM-SC and SM-NM interfaces with amplitudes $\tilde{t}_{\rm SC}$ and $\tilde{t}_{\rm NM}$, respectively. Explicitly, the coupling terms  can be written as
\begin{align}
H^{SM-SC}=\sum_{\bm i_1, \bm j_1, \sigma}\left[\tilde{t}_{SC} c_{\bm i_1 \sigma}^\dagger a_{\bm j_1\sigma} + h.c.\right], \ \label{Eq6} \\
H^{SM-NM}=\sum_{\bm i_2, \bm j_2, \sigma}\left[\tilde{t}_{NM} c_{\bm i_2 \sigma}^\dagger b_{\bm j_2\sigma} + h.c.\right],  \label{Eq7}
\end{align}
where $\bm i_1=(i_x, i_y, i_{1z})$ and $\bm j_1=(j_x, j_y, j_{1z})$ label lattice sites near the SM-SC interface inside the SM and SC regions, respectively,  $\bm i_2$ and $\bm j_2$ are neighboring sites across the SM-NM interface, while $\tilde{t}_{SC}$ and $\tilde{t}_{NM}$ are the hopping matrix elements that characterize the couplings across the two interfaces.

The effective low-energy theory for the proximity effects at the SM-SC and SM-NM interfaces can be obtained using a Green's function formalism by integrating out the SC and NM degrees of freedom. The effective Green's function has the form
\begin{equation}
G_{ij}(\omega) =\left[\omega \delta_{ij} - H_{ij}^{\rm SM} - H_{ij}^{\rm V} -\Sigma_{ij}^{\rm SC}(\omega)-\Sigma_{ij}^{\rm NM}(\omega)\right]^{-1}.                  \label{Geff}
\end{equation}
Here $i$ and $j$ are spatial indices and the proximity-induced SM self-energies are proportional to the surface Green's functions of the SC and NM, respectively.\cite{Stanescu2013b} For example, integrating out the SC degrees of freedom results in a local self-energy contribution to the SM Green's function,
\begin{equation}
\Sigma_{{\bm i}_1, {\bm i}_1^\prime}^{SC}(\omega) = |\tilde{t}_{SC}|^2 G_{SC} (\omega, {\bm j}_1, {\bm j}_1^\prime),  \label{Eq8}
\end{equation}
where $G_{SC}$ is the Green's function of the superconductor, which contains both normal and anomalous contributions, and $\tilde{t}_{SC}$ is the hopping across the SM-SC interface between the pairs of sites $(i_1, j_1)$ and $(i_1^\prime, j_1^\prime)$. Next, we assume that the SC is described by the Hamiltonian from Eq. (\ref{Eq4}) and calculate the Green's function $G_{SC}$. A major simplification results from the observation that $G_{SC} (\omega, {\bm j}_1, {\bm j}_1^\prime)= G_{SC} (\omega, |{\bm j}_1- {\bm j}_1^\prime|)$ is a short-range function of $|{\bm j}_1- {\bm j}_1^\prime|$ that decays much faster than $1/k_{SM}$, where $k_{SM}$ is some characteristic wave vector of the SM (e.g., the largest Fermi $k$ vector in a multiband system). Consequently, we can use the local approximation  $G_{SC} (\omega, |{\bm j}_1- {\bm j}_1^\prime|)\approx G_{SC} (\omega, 0)$. Within this approximation, the self-energy (\ref{Eq8}) induced by the superconductor will correspond to a local contribution acting at the SM-SC interface. Similar considerations hold for the self-energy that results by integrating out the normal-metal degrees of freedom. 
In the relevant long-wavelength low-energy regime of interest, these self-energies generate local contributions at the SM-SC and SM-NM interfaces having the explicit forms
\begin{align}
\!\Sigma_{ij}^{\rm SC} (\omega)\!&\!=\! - \! \delta_{ij}\delta_{i_z, 0}\!\left(\!\gamma_{\rm SC}\frac{\omega \!+\!\sigma_y \tau_y \Delta_0}{\sqrt{\Delta_0^2\!-\!\omega^2}} \!+\! \gamma_{\rm SC}\frac{2|t_{\rm SC}|\!-\!\mu_{\rm SC}}{2t_{\rm SC}^2\nu(\mu_{\rm SC})}\!\right)\! \label{SigSC} \\
\!\Sigma_{ij}^{\rm NM} (\omega)\!&\!= -\delta_{ij}\delta_{i_z, 0}\left(i~\gamma_{\rm NM}+\gamma_{\rm NM}\frac{2|t_{\rm NM}|-\mu_{\rm NM}}{2t_{\rm NM}^2\nu(\mu_{\rm NM})}\right), \label{SigNM}
\end{align}
\noindent where $\mu_{\rm SC}$ and $\mu_{\rm NM}$ are the Fermi energies of SC and NM, respectively, and  $\sigma_i$ and $\tau_i$ are Pauli matrices in the spin and particle-hole spaces, respectively. The effective couplings $\gamma_{\rm SC}$ and
$\gamma_{\rm NM}$ can be expressed as $\gamma_{\rm x}= \nu(\mu_{\rm x})\tilde{t}_{\rm x}^2/2$, where ${\rm x}$ represents the SC or the NM,
in terms of the surface density of states at the Fermi level, $\nu(\mu_{\rm x})= \sqrt{1 - (2t_{\rm x}-\mu_{\rm x})^2/4t_{\rm x}^2}/|t_{\rm x}|$, and
the tunneling amplitudes at the interfaces, $\tilde{t}_{\rm SC}$ and $\tilde{t}_{\rm NM}$.\cite{Stanescu2013b}
We note that Eq. (\ref{SigSC}) holds for energies within the bulk SC gap, i.e., for $-\Delta_0 <\omega <\Delta_0$.  We also note that for an arbitrary strength of the SM-SC coupling, $\gamma_{SC}$, the induced gap does not have a simple analytic expression and, moreover, depends nontrivially on the thickness of the SM wire.\cite{Stanescu2013a} However, in the weak-coupling limit characterized by $\gamma_{SC} \ll \Delta_0$, where $\Delta_0$ is the bulk SC gap, and  $\gamma_{SC}$ is much smaller than the interband spacing, we have
\begin{equation}
\Delta_{ind}= \frac{\gamma_{SC} \Delta_0}{\gamma_{SC} + \Delta_0}.  \label{Eq9}
\end{equation}
We also note that, in the most general case, the effective SM-SC coupling can be different for different SM bands and that proximity-mediated interband couplings are possible. These refinements are not considered in the present work, which focuses on the effects of the SM-NM coupling rather than the proximity effect induced at the SM-SC interface, which has already been extensively studied in the literature.

To calculate the local density of states (LDOS) inside the SM wire, we determine the SM Green's function matrix in Eq. (2), $G_{ij}(\omega)=\left[\omega \delta_{ij} - H_{ij}^{\rm SM} - H_{ij}^{\rm V} -\Sigma_{ij}^{\rm SC}(\omega)-\Sigma_{ij}^{\rm NM}(\omega)\right]^{-1}$, by numerically performing the matrix inversion for $H_{ij}^{\rm SM}$ and $H_{ij}^{\rm V}$ corresponding to Eqs. (\ref{Eq2}) and (\ref{Eq3}), respectively, and the self-energies given by Eqs. (3) and (4).  If ${\cal R}$ is a certain region of the wire (e.g., the barrier region, or the segment covered by the SC), the LDOS integrated over ${\cal R}$ (notation $DOS_{\cal R}$) is given by
\begin{equation}
\rho_{\cal R}(\omega) = -\frac{1}{\pi} \sum_{i \in {\cal R}} {\rm Im}[G_{ii}(\omega)].  \label{Eq10}
\end{equation}

To test the accuracy of the effective theory given by Eqs. (\ref{Geff}), (\ref{SigSC}), and (\ref{SigNM}), we have calculated the LDOS integrated over the SM volume for the SC-SM-NM layered structure shown schematically in the inset of Fig. \ref{FFig2}a and compared the results with the LDOS obtained by exactly diagonalizing the full Hamiltonian (which is straightforward in this geometry). The excellent agreement between the two calculations is shown in Fig. \ref{FFig2}(d).

\subsection{Differential conductance}

\begin{figure}[tb]
\begin{center}
\includegraphics[width=0.48\textwidth]{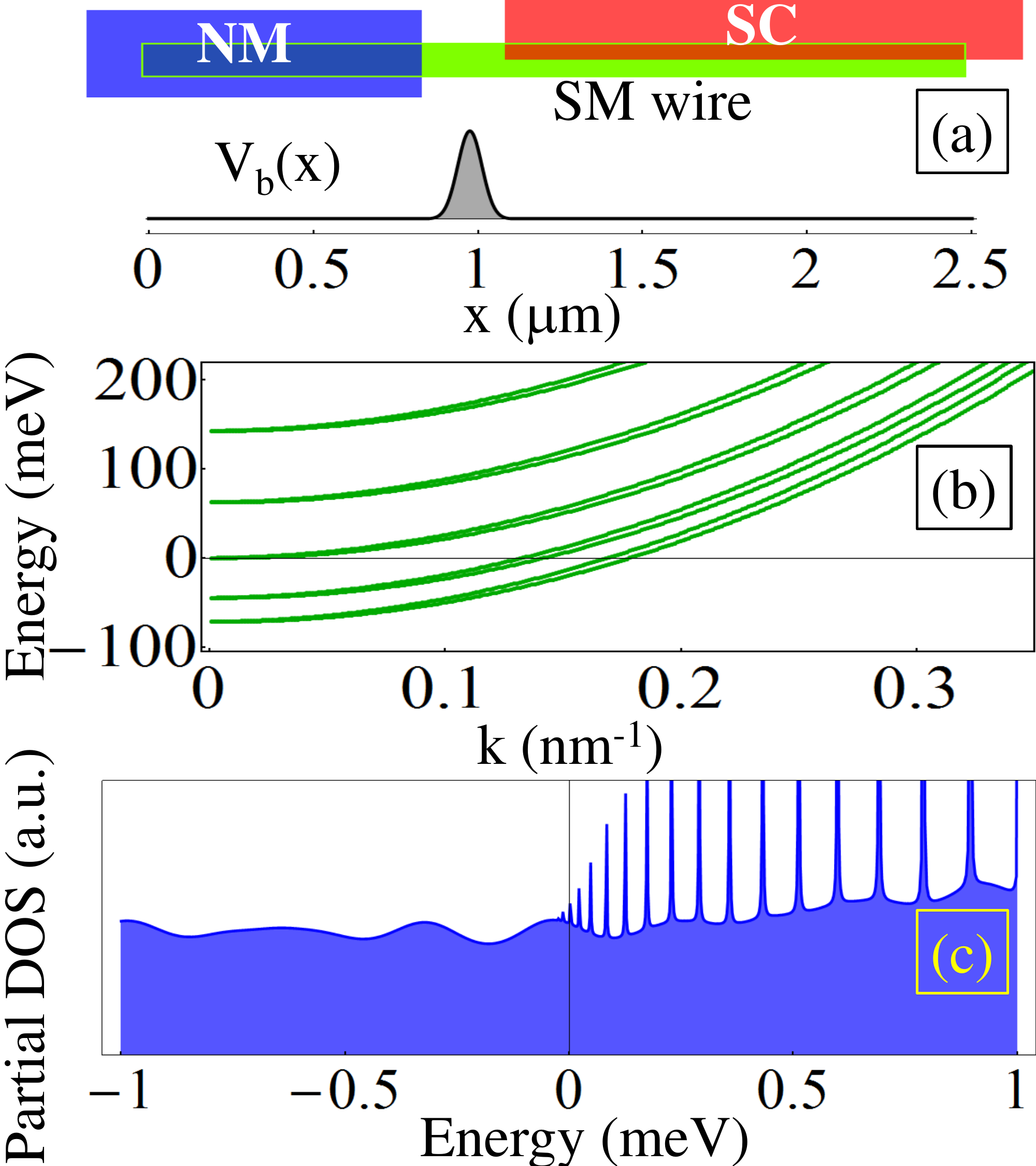}
\vspace{-8mm}
\end{center}
\caption{(Color online)  (a) Schematic representation of a SM Majorana wire proximity coupled to a NM and a SC (top view). A barrier potential $V_b(x)$ is applied in the region of the wire that is not covered by the NM or the SC. (b) Spectrum of the SM wire in the absence of coupling ($\gamma_{SC(NM)}=0$). The chemical potential is placed near the bottom of the third band. (c) Local density of states integrated over the barrier region (i.e., around $x=$1 $\mu$m) for a wire coupled to the NM (only). The smooth background is due to states from the low-energy bands that penetrate through the barrier and hybridize strongly with NM states. The peaks correspond to states from the top occupied band that are confined to the right side of the barrier and hybridize very weakly with the NM. }
\vspace{-4mm}
\label{FFig3}
\end{figure}

The differential conductance calculations are performed using a simplified model consisting of two coupled chains.
We note that implementing this technique using the more sophisticated model described by Eqs. (\ref{Htot}-\ref{Eq7}) is straightforward, but involves substantial numerical costs that are not justified by our main goal (i.e., to show that the manifestations of the soft-gap phenomenon in the LDOS and in $dI/dV$ are related and stem from the same source).
Specifically, the SM wire is modeled using a Hamiltonian similar to that from Eq. (\ref{Eq2}), but having the proximity-induced SC gap already included at the mean-field level:
\begin{align}
H^{\rm SM} = -\sum_{{\bm i}, {\bm \delta},\sigma}  \left[ t_{SM}c_{{\bm i}\!+\!{\delta_x}\sigma}^{\dagger}c_{{\bm i}\sigma}+  t_{SM}^{\perp}c_{{\bm i}\!+\!{\delta_y}\sigma}^{\dagger}c_{{\bm i}\sigma}\right]  \nonumber \\
 -\mu \sum_{{\bm i}, \sigma} c_{{\bm i}\sigma}^{\dagger}c_{{\bm i}\sigma}  \nonumber
+\Gamma \sum_{{\bm i}}[c_{{\bm i}\uparrow}^{\dagger}c_{{\bm i}\downarrow} + h.c.]~~~~~ \\
+ \frac{i}{2}\sum_{{\bm i},{\bm \delta}}\left[\alpha c_{{\bm i}+{\bm \delta}_x}^{\dagger}{\sigma}_y c_{{\bm i}} -  \alpha^{\perp} c_{{\bm i}+{\bm \delta}_y}^{\dagger}{\sigma}_x c_{{\bm i}} + {\rm h.c.} \right],  \label{Eq11} \\
+ \Delta_{ind}\sum_{\bm i}^{(i_x>L_N)} (c_{{\bm i}\uparrow}^\dagger c_{{\bm i}\downarrow}^\dagger + c_{{\bm i}\downarrow}c_{{\bm i}\uparrow}), ~~~~~~~~~ \nonumber
\end{align}
where ${\bm i} = (i_x, i_y)$ with $i_y=1, 2$  gives the position along the two chains, $(t_{SM}, t_{SM}^{\perp})$ and $(\alpha, \alpha^{\perp})$ are the intra-chain and and inter-chain hoppings and Rashba coefficients, respectively, $\mu$ is the chemical potential, and  $\Gamma$ is the Zeeman field. The normal segment of the double chain corresponds to the first $L_N$ sites, while the rest of the double chain is coupled to the SC and, as a result, has an induced pairing potential $\Delta_{ind}$.  Note that including the pair potential directly into the SM Hamiltonian corresponds to the static approximation $\sqrt{\Delta_0^2-\omega^2}\approx \Delta_0$ in Eq. (\ref{SigSC}) for the self-energy (see, for example, Ref. \onlinecite{Stanescu2013b} concerning the accuracy of this approximation).
 A potential barrier given by Eq. (\ref{Eq3}) is applied in the region that separates the normal and superconducting segments [see Fig. \ref{FFig3}(a)]. The normal segment of the wire is in contact with a metallic lead also described by a double-chain model corresponding to Hamiltonian (\ref{Eq5}).  The parameters used in the calculation correspond to an effective mass $m_{eff}^{SM} = 0.016m_0$ for the semiconductor wire and $m_{eff}^{NM} = 0.4 m_0$ for the metallic lead. The interchain coupling $t_{SM}^{\perp}=4$ meV corresponds to a splitting of $8$ meV between the two SM bands. The Rashba coefficients are $\alpha a=200.0$ meV \AA and $\alpha^{\perp}=0$.
Following the standard scattering matrix formalism, we solve the eigenvalue problem for the full tight-binding Bogoliubov-de Gennes (BdG) Hamiltonian with open boundary conditions and determine the normal and anomalous reflection amplitudes. More specifically, let us assume that the transverse modes corresponding to the multichain problem are described by the wave functions $\Phi_n(i_y)$, where for two chains $n$ takes two values corresponding to the symmetric and antisymmetric modes, and consider an incoming electron with spin $\sigma$ in channel $n$. The wave function of energy $E$ is a four-component spinor with values on the first two sites of the lead given by
\begin{align}
\Psi_{i_x=0}^{(n,\sigma)} = \left(
\begin{array}{c}
\delta_{\sigma\uparrow} \\
\delta_{\sigma\downarrow} \\
0 \\
0
\end{array}\right)\Phi_n
+\sum_{n^\prime}\left(
\begin{array}{c}
r^N_{n n^\prime, \sigma\uparrow} \\
r^N_{n n^\prime, \sigma\downarrow} \\
r^A_{n n^\prime, \sigma\uparrow} \\
r^A_{n n^\prime, \sigma\downarrow}
\end{array}\right)\Phi_n^\prime   \label{Eq12}
\end{align}
and
\begin{align}
\Psi_{i_x=1}^{(n,\sigma)} = \left(
\begin{array}{c}
\delta_{\sigma\uparrow} \\
\delta_{\sigma\downarrow} \\
0 \\
0
\end{array}\right)\Phi_n e^{ik_e^n a}
+\sum_{n^\prime}\left(
\begin{array}{c}
r^N_{n n^\prime, \sigma\uparrow} \\
r^N_{n n^\prime, \sigma\downarrow} \\
0 \\
0
\end{array}\right)\Phi_n^\prime  e^{-ik_e^{n^\prime} a}  \nonumber \\
+\sum_{n^\prime}\left(
\begin{array}{c}
0 \\
0 \\
r^A_{n n^\prime, \sigma\uparrow} \\
r^A_{n n^\prime, \sigma\downarrow}
\end{array}\right)\Phi_n^\prime e^{ik_h^{n^\prime} a}, ~~~~~~~~~~~~~~ \label{Eq13}
\end{align}
where $r^N(E)$ and $r^A(E)$ represent normal and anomalous reflection coefficients corresponding to different incoming and outgoing channels. The  electron and hole wave vectors corresponding to an eigenstate with energy $E$ satisfy the equations $E_{NM}^n(k_e^n)-\mu_{NM} = E$ and $E_{NM}^n(k_h^n)-\mu_{NM} = -E$, respectively. Here $E_{NM}^n(k)$ is the energy of the normal lead corresponding to the transverse mode $n$.
After solving the eigenvalue problem numerically and determining the reflection coefficients over the relevant energy range, we calculate the zero-temperature differential conductance, $G_0(V) = \frac{dI}{dV}$, using the anomalous reflection amplitudes,
\begin{equation}
\frac{dI}{dV} = \frac{2e^2}{h} \sum_{n, n^\prime}\sum_{\sigma\sigma^\prime} |r^A_{n n^\prime, \sigma\sigma^\prime}(V)|^2.  \label{Eq14}
\end{equation}
The finite temperature conductance can be obtained by convolving $G_0$ with the derivative of the Fermi function,
\begin{equation}
G(V, T) = \int d\omega G_0(\omega) \frac{1}{4 T \cosh^2[(V-\omega)/2T]}.   \label{Eq15}
\end{equation}

\section{Results}\label{results}

\subsection{Emergence of the soft-gap in clean semiconductor-superconductor hybrid structures}

Consider a SC-SM-NM hybrid structure consisting of a SM nanowire proximity coupled to an $s$-wave SC and a normal-metal lead, as shown schematically in Fig. \ref{FFig3}(a). The structure is modeled by the Hamiltonian given by Eq. (\ref{Htot}) and the partial density of states obtained by integrating the LDOS over the barrier region (i.e., the short nanowire segment that is not covered by the SC or the NM) is determined using the effective theory described in the previous section.
First, we focus on the proximity effect due to the coupling to the normal lead and set $\tilde{t}_{SC}=0$. We assume that the SM wire has several occupied subbands in the absence of the lead, as shown in Fig.~\ref{FFig3}(b). The dependence of the LDOS on the {bias potential} for a nonzero coupling to the lead, $\gamma_{NM}=0.2$ meV, is shown in Fig.~\ref{FFig3}(c). Note that the confining barrier potential $V_b$ affects very differently the  states corresponding to different transverse subbands.  The smooth background in Fig.~\ref{FFig3}(c) is due to states from the low-energy SM bands that penetrate through the barrier and hybridize strongly with the normal metal, whereas the peaks originate from the top occupied subband and correspond to states that are confined to the right side of the barrier and hybridize very weakly with the NM. We note that increasing the strength of the potential barrier has two effects. (i) The weight of the sharp peaks decreases as the states from the topmost occupied band are pushed out of the barrier region. Eventually, these peaks are no longer observable. (ii) The modulation of the smooth background increases and, eventually, this background develops into a discrete set of peaks corresponding to states from the low-energy occupied bands. The peaks associated with a particular band emerge when the corresponding states are confined by the barrier potential within the SC  segment of the wire. The broadening induced by the imaginary part of $\Sigma^{\rm NM}$ from Eq. (\ref{SigNM}) becomes negligible, as these states have exponentially  vanishing amplitudes outside the SC region, more specifically at the SM-NM interface.

\begin{figure}[tb]
\begin{center}
\includegraphics[width=0.48\textwidth]{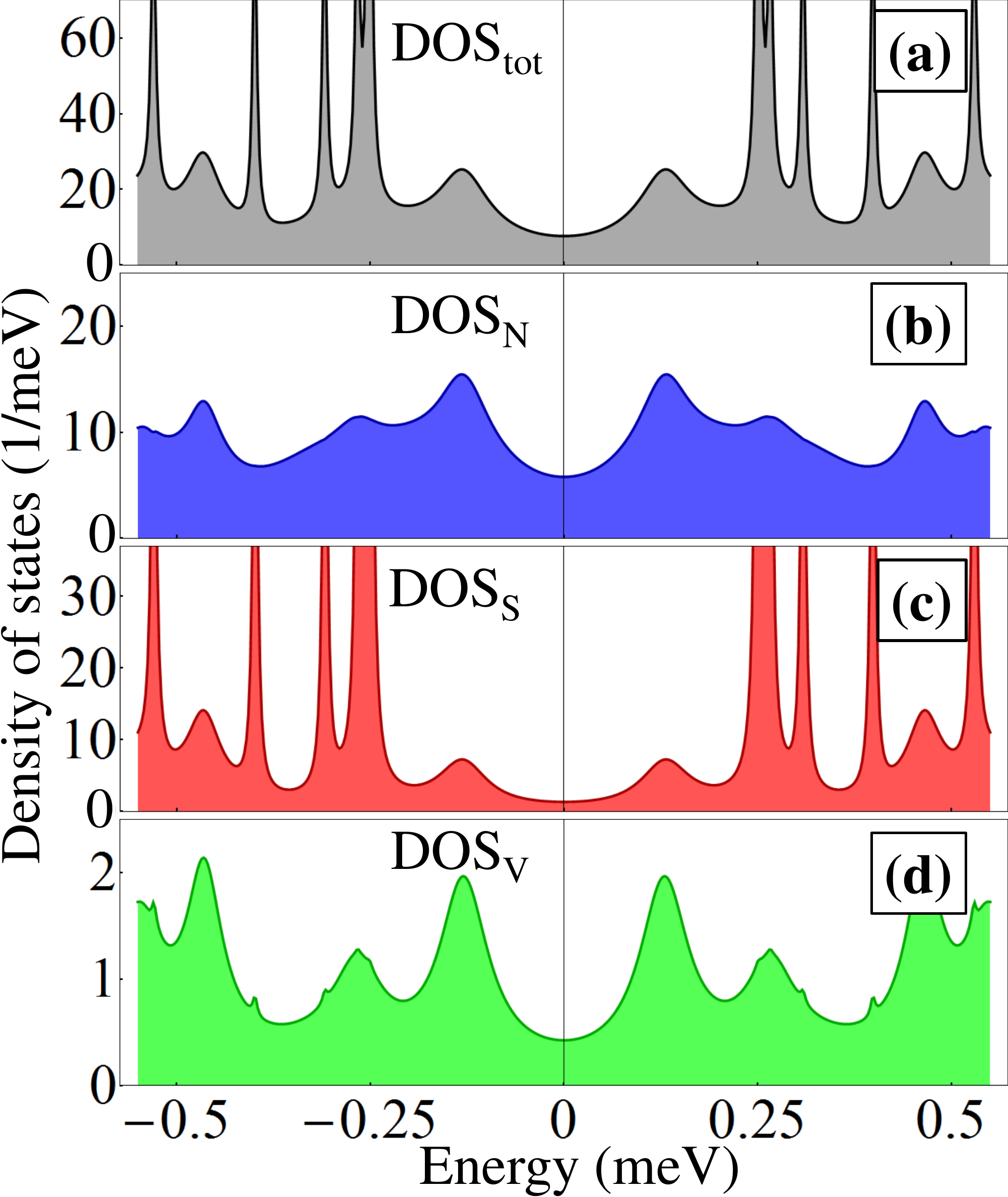}
\vspace{-8mm}
\end{center}
\caption{(Color online) Density of states (inside the SM wire)  at zero Zeeman field in the presence of both NM and SC couplings. (a) Total DOS inside the SM wire. (b) LDOS integrated over the region covered by the NM contact. (c) LDOS integrated over the SC region. (d) LDOS integrated over the barrier region. Note that the {sharp} peaks are due states that are confined to the SC region (and penetrate slightly inside the barrier region). Parameters: $\gamma_{SC}=0.5$ meV, $\gamma_{NM}=0.1$ meV, $\Gamma=0$, and $V_b^{max}=5$ meV.}
\vspace{-4mm}
\label{FFig4}
\end{figure}

Next, we discuss the quasiparticle DOS in the presence of both couplings, $\tilde{t}_{\rm SC}, \tilde{t}_{\rm NM} \neq 0$, at $\Gamma=0$. Note that  the soft-gap is a qualitative feature of the multiband NM-SM-SC system that is independent of the applied magnetic field. The results for the (total) DOS in nanowire are shown in Fig.\ref{FFig4}(a). Notice the substantial background contribution. To identify its origin, it is instructive to calculate the local density of states integrated over the relevant segments of the wire. Specifically, we divide the nanowire into three regions: (1) the segment of the nanowire covered by the normal lead, (2) the portion covered by the superconductor, and (3) the uncovered portion of the nanowire where the confining gate potential is applied. The contributions to the DOS from regions (1)-(3) are shown in Figs. \ref{FFig4}(b)-\ref{FFig4}(d), respectively. Similar to the normal
case shown in Fig. \ref{FFig3}(c), the density of states in the wire consists of a superposition of smooth contributions given either by delocalized states that span the whole length of the nanowire or by states that are confined inside the normal region and sharp peaks emerging from the states confined inside the superconducting segment. Note that the states responsible for the sharp features correspond to the topmost occupied band (the Majorana band), which are confined inside the SC segment and, consequently, hybridize very weakly with the normal lead.

\subsection{Comparison between LDOS and transport soft-gap features}

\begin{figure}[t]
\begin{center}
\includegraphics[width=0.48\textwidth]{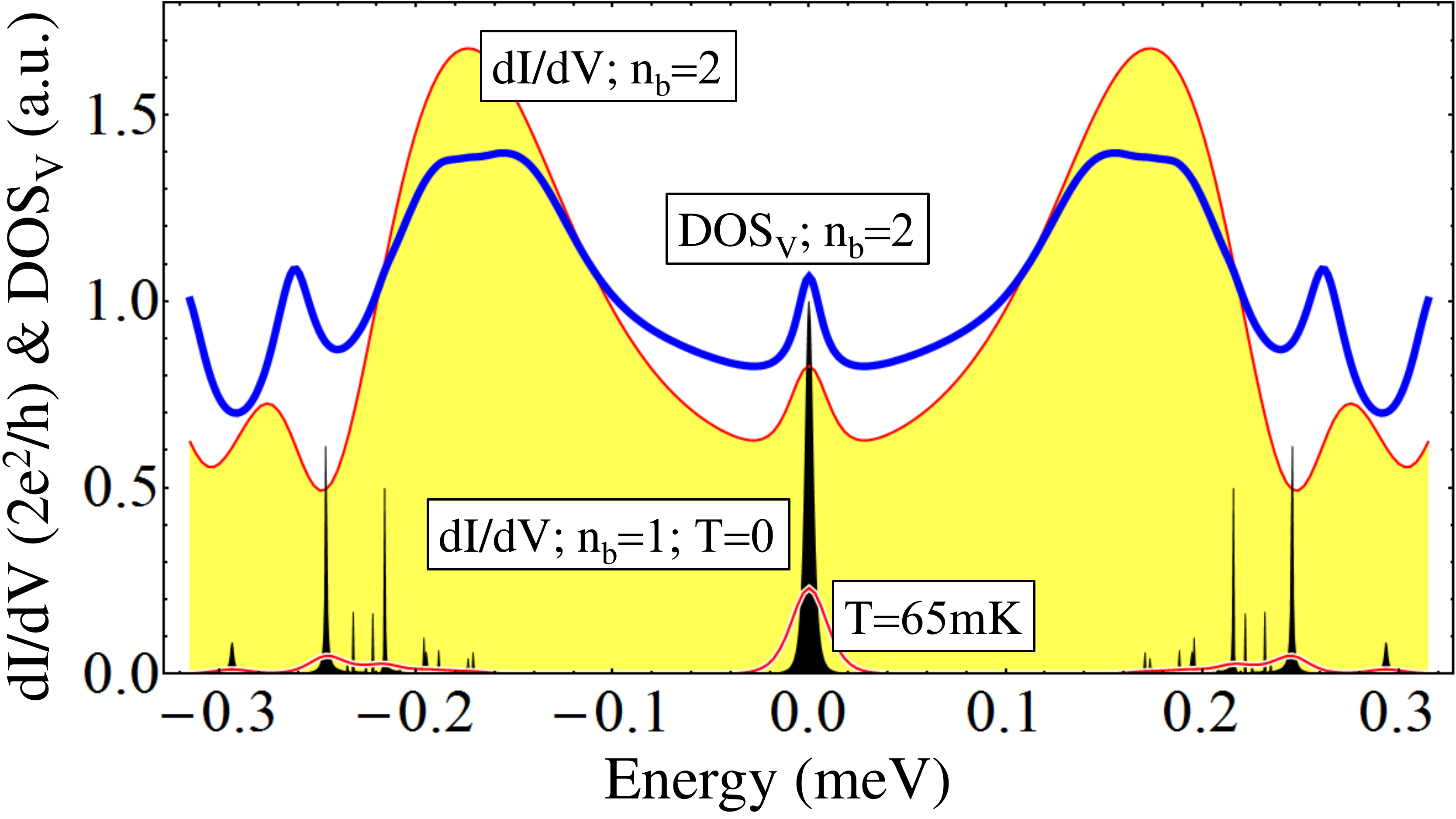}
\vspace{-7mm}
\end{center}
\caption{(Color online)  Differential conductance and LDOS (integrated over the barrier region -- $DOS_V$) calculated using the double-chain model. The sharp features represent the conductance of a system with a single occupied band ($n_b=1$) at zero temperature, $T=0$ (filled black lines). Notice the quantization of the Majorana-induced zero-bias peak. At finite temperature ($T=65$ mK), the sharp features (including the Majorana peak) become suppressed (red line). No background that could be associated with the soft-gap phenomenon exists for $n_b=1$. $DOS_V$ (not shown) has a similar structure. When two SM bands are occupied ($n_b=2$), both $dI/dV$ (red line with yellow filling) and $DOS_V$ (blue line) show a zero-bias peak on top of a substantial smooth background. The Majorana peak is strongly suppressed at  finite temperature (here $T=65$ mK), but the background is practically independent of $T$. In all these calculations the Zeeman field is $\Gamma=0.4$ meV.}
\vspace{-0mm}
\label{FFig5}
\end{figure}

As we have argued above, the soft-gap phenomenon is due to the band-selective coupling between the SM states and the metallic lead and can be understood as a band-dependent combination of metallic and tunnel junction behavior. The soft-gap behavior manifests itself in transport (as a  substantial subgap contribution to the differential conductance), as well as in the local density of states. In other words, the critical ingredient is the presence of a metallic lead coupled to the nanowire and not the existence of an actual, nonzero charge current in the system. To firmly establish these points, we calculate both $dI/dV$ and the LDOS using the double-chain model for two different cases: (a) single-band occupancy (the chemical potential $\mu_{SM}$ is positioned near the bottom of the lowest energy band) and (b) two-band occupancy ($\mu_{SM}$ near the bottom of the second band). The results are shown in Fig. \ref{FFig5}. These results clearly support three main conclusions. (1) The soft-gap requires multiband occupancy. Since many of the previous model calculations were done in the one-dimensional limit (i.e., on a single chain), they were not able to capture this phenomenon. (2) The soft gap observed in the tunneling conductance, $dI/dV$, and in the LDOS integrated over the barrier region, $DOS_V$, are two manifestations of the same phenomenon, i.e., the result of the strong coupling to the normal lead. (3) The (intrinsic) soft-gap can be clearly eliminated by either realizing the single-band occupancy limit, or by completely  eliminating the metallic lead (i.e., using a probe other than transport to observe the Majorana bound states). A third possibility involving reduced occupancy in combination with an optimized  barrier potential will be discussed below.

\subsection{Controlling the soft-gap feature}

\begin{figure}[tb]
\begin{center}
\includegraphics[width=0.48\textwidth]{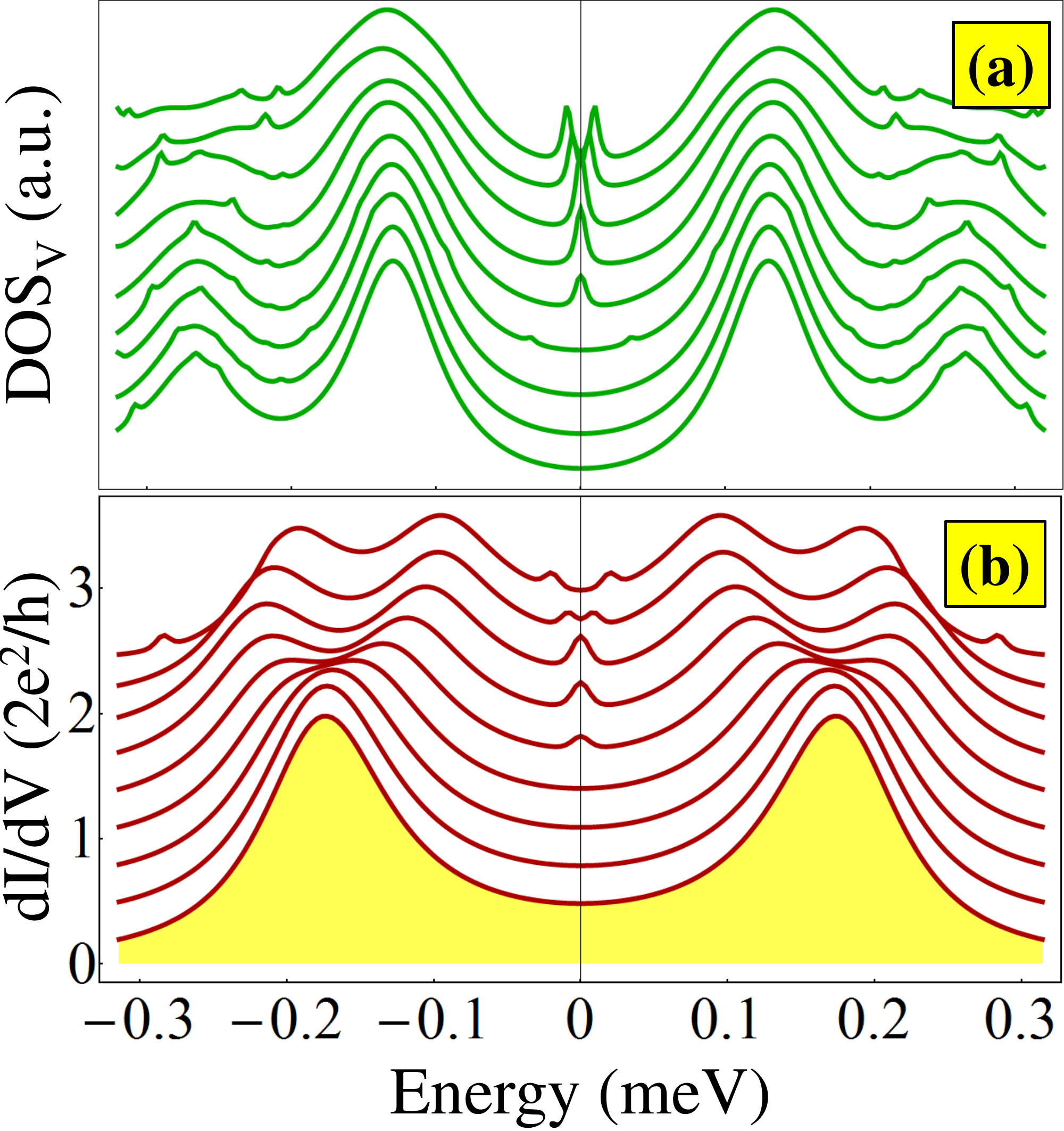}
\vspace{-6mm}
\end{center}
\caption{(Color online) (a) Magnetic field dependence of the LDOS integrated over the barrier region for a system described by Eqs. (\ref{Geff}), (\ref{SigSC}), and  (\ref{SigNM}) with $\gamma_{NM}=0.1$ meV,   $\gamma_{SC}=0.5$ meV, and $L_S=$1.1 $\mu$m. (b) Differential conductance for a double-chain wire with
$\Delta_{ind}=0.25$ meV, $L_S=$1.1 $\mu$m, and $T=0.65$ mK. The curves (shifted for clarity) correspond to Zeeman fields from 0 (bottom) to 0.8 meV (top).  A Majorana peak emerges above a certain critical field and splits at high Zeeman fields.  Note the close correspondence between the two types of calculations which involve systems described by different models and having different parameters.}
\vspace{-4mm}
\label{FFig6}
\end{figure}

We now consider the dependence of the soft-gap feature on the applied magnetic field and the potential barrier.
To understand the experimental measurements involving tunneling spectroscopy, it is instructive to compare the magnetic field dependence of the LDOS in region (3) (the barrier region) and the field dependence of the differential conductance. The results are shown in Fig.~\ref{FFig6}. In both cases one can notice a substantial subgap background and the emergence of a zero-bias peak (ZBP) above a certain critical value of the magnetic field. This peak is due to a Majorana mode localized at the left end of region (2) and leaking out into the barrier region. The splitting of the ZBP at large values of the Zeeman field (see Fig.~\ref{FFig6}) is due to the overlap of the wave functions corresponding to the Majorana bound states localized at the opposite ends of the superconducting segment of the wire.\cite{Meng_splitting, DSarma2012}
One can notice the close similarity between the qualitative features characterizing the two quantities.

\begin{figure}[tb]
\begin{center}
\includegraphics[width=0.5\textwidth]{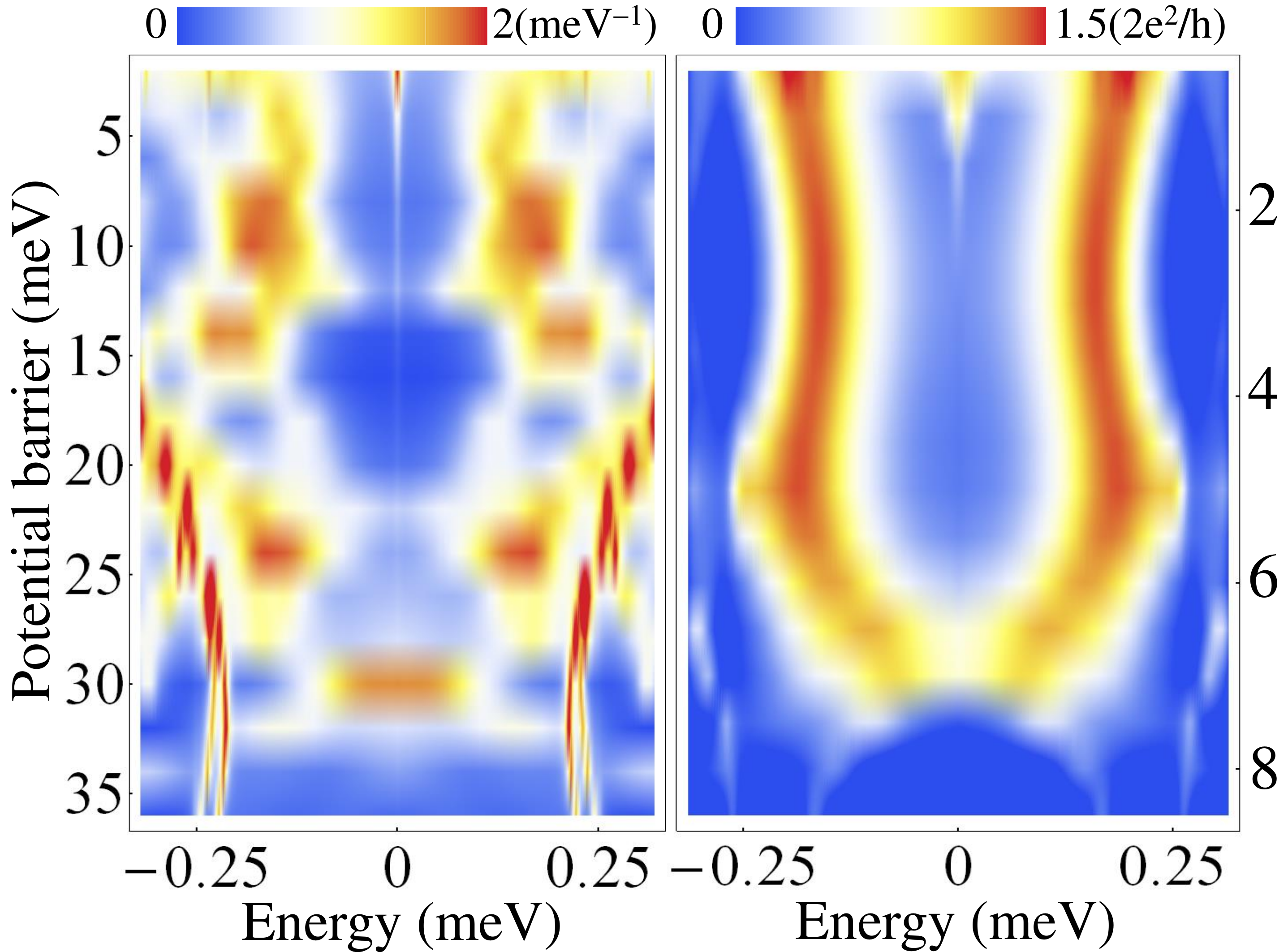}
\vspace{-3mm}
\end{center}
\caption{(Color online) {Dependence of the LDOS integrated over the barrier region (left) and of $dI/dV$ (right) on the barrier height $V_b^{\rm max}$.  Note that increasing the barrier height reduces the amplitude of the zero-bias peak as the Majorana bound state penetrates less inside the barrier region and decouples from the lead.  In addition, varying the barrier potential changes the energy of the delocalized states that extend throughout the wire and, as a result, modifies the profile of the soft-gap. The finite-energy sharp features present near the induced SC gap edge ($\Delta_{\rm ind} \approx 0.25$ meV, left panel) are due to the states from the second highest occupied band that become confined inside the superconducting region. On the other hand, the smooth low-energy maximum corresponding to
 $V_b^{\rm max}=30$ meV is generated by a state from the second highest occupied band that crosses the chemical potential and is confined inside the normal segment of the wire. The inter-band spacings are $35$ meV (left) and $8$ meV (right) and the Zeeman field is $\Gamma=0.4$ meV. }}
\vspace{-3mm}
\label{FFig7}
\end{figure}

The effective coupling between the normal-metal lead and the active system can, in principle, be controlled through the gate-tunable potential barrier.\cite{Mourik2012, Das2012, Deng2012, Fink2012, Churchill2013} The dependence of $DOS_V$ and $dI/dV$ on the height of the barrier potential for a Gaussian potential with $w_b=52$nm is shown in Fig.~\ref{FFig7}. One can notice that the barrier affects the subgap states and the  Majorana zero-energy state differently. Indeed, let us consider the high barrier limit (i.e., barrier height larger than the chemical potential in the SM nanowire) and study the evolution of the subgap density of states. Upon a decrease of the barrier height, the soft-gap feature is emerging first, and then the zero-energy peak becomes visible. This fact confirms that the soft-gap originates from the low-lying subbands that can more easily penetrate through the barrier and thus become strongly hybridized with the metallic lead. By contrast, the Majorana zero-energy peak originates from the weakly coupled topmost band. We also find that the change of the potential barrier modifies the energies of the extended states that contribute to the soft-gap and, consequently, modifies its profile. We conclude that the structures used in current tunneling experiments\cite{Mourik2012, Das2012, Deng2012, Fink2012, Churchill2013} are not optimally designed for suppressing the soft-gap feature (e.g. by increasing the barrier potential height)  while maintaining the visibility of the Majorana zero-bias peak. We identify the main problem as the broad barrier profile that characterizes these structures.

\begin{figure}[t]
\begin{center}
\includegraphics[width=0.48\textwidth]{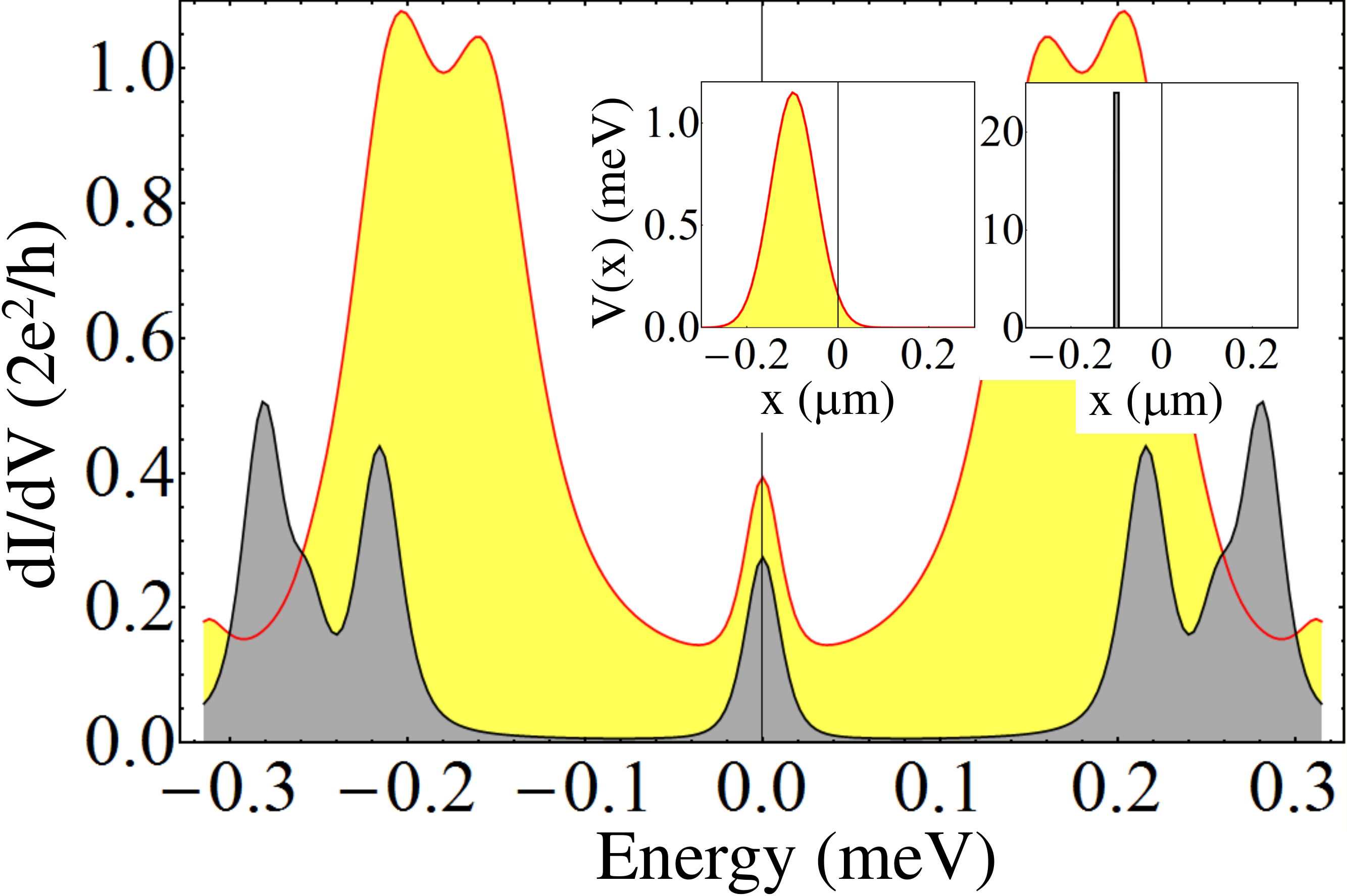}
\vspace{-3mm}
\end{center}
\caption{(Color online) Comparison between the differential conductance characterizing two hybrid systems with $\Delta_{ind}=0.25$ meV, $L_S=$2 $\mu$m, $\Gamma=$0.4 meV and different potential barriers (see inset). The two barriers have the same transparency for the Majorana band, but significantly different transparencies for the low-energy occupied band.  This results in the two Majorana ZBPs having the same weight and the soft-gap being strongly suppressed in the system with a narrow barrier.}
\vspace{-0mm}
\label{FFig8}
\end{figure}

Finally, we address a question of significant practical importance: if merely increasing the height of the potential barrier is likely to generate the disappearance of the ZBP before producing a hard gap, what should be done to ensure the coexistence of  both the ZBP and the hard gap? The solution involves two basic requirements: i) reducing the occupancy of the nanowire and ii) generating a very narrow barrier. Of course, this assumes that all extrinsic effects are already eliminated by carefully engineering clean hybrid structures. The first requirement is less restrictive than the single-band occupancy condition (if this can be realized, the soft-gap feature is automatically eliminated). The rationale behind both the first and the second requirement is to minimize the ratio between the transparencies corresponding to the lowest-energy band and the topmost band. These could still be significantly different, but this difference should not be more than about an order of magnitude. For example, a transparency for the Majorana band $T_M=0.01$ (which corresponds to $Z_M\approx 10$ for a $\delta$-function-type barrier) ensures the visibility of the Majorana ZBP. If the lowest-energy band has a transparency $T_1=10T_M$ (which corresponds to $Z_1=3$), it will still be in the weak-coupling limit and will generate an extremely weak in-gap background. Satisfying these conditions using a broad barrier may be extremely difficult (it requires very small inter-band gaps, i.e., a thick wire with just a couple of occupied bands). On the other hand, using a narrow barrier facilitates the realization of these conditions. In fact, we believe that the simplifying assumption of a barrier that can be modeled by a single-site/$\delta$-function-like potential has prevented the observation of soft-gap features in previous multiband model calculations.\cite{PradaPRB'12,PientkaPRL'12,RainisPRB'13}
To illustrate the major difference between the effect of broad and narrow
barriers, we calculate the differential conductance for two systems having the same parameters, except for the barrier potentials, which have different profiles, as shown in the insets of Fig. \ref{FFig8}. The heights of the two barriers were adjusted so that the two systems have the same transparency for the Majorana band. This translates into the two Majorana ZBPs having the same weight (note that the ZBP corresponding to the wide barrier sits on top of the smooth subgap background generated by states from the lowest-energy band). The strong suppression of the soft-gap feature in the system with a narrow barrier is due to the fact that the transparency of the low-energy band (with a Fermi energy of about $8$ meV) is much smaller than the transparency of the same band in the wide barrier system. This can be easily understood by comparing the Fermi energy with the heights of the two barriers.

\section{Conclusions}

We have developed a comprehensive theory for  the NM-SM-SC hybrid heterostructures using realistic models for the semiconductor nanowire, the normal-metal lead, and the $s$-wave superconductor. We have identified a distinct physical mechanism, namely an ``inverse proximity'' effect by the normal metal on the semiconductor nanowire at the semiconductor-metal interface, which by itself generically leads to soft-gap features in the multiband wire system by virtue of producing substantial subgap conductance.  This inverse proximity effect is, in principle, always present, and has a strength that depends on the details of the tunnel barrier at the semiconductor--normal-metal junction and on the number of conducting subbands active in the nanowire. Our model, without including any external effects such as interface disorder, captures the phenomenology observed in recent Majorana experiments\cite{Mourik2012, Das2012, Deng2012, Fink2012, Churchill2013} and allows one to explain the so-called soft-gap issue. We show that the substantial subgap conductance in these experiments originates from the multiple subbands in the nanowire having vastly different transmission probabilities through the barrier defined by the gate. We emphasize that this is an intrinsic contribution to the soft-gap issue; the extrinsic effects such as interface disorder will further exacerbate this problem.  Thus our results have important implications for the ongoing Majorana experiments where the soft-gap issue is the main roadblock to braiding and performing quantum computation with Majorana modes.  We show that, to interpret the experimental results,\cite{Mourik2012, Das2012, Deng2012, Fink2012, Churchill2013} one has to consider different coupling of the occupied subbands in the nanowire to the normal-metal lead. By calculating local density of states and differential tunneling conductance, we show that the ``Majorana band'' (i.e., highest subband)  has generically the weakest coupling to the lead. Therefore, one has to lower the barrier in order to increase signal-to-noise ratio which, in turn, increases the normal-metal coupling for the lowest occupied subbands leading to a significant hybridization of these states with the lead (i.e., the soft-gap problem).
To reduce the subgap background, we suggest decreasing the NM-nanowire coupling and the length of the normal nanowire segment, realizing narrow barriers, or using alternative, less invasive detection schemes involving an STM\cite{ZeroBiasAnomaly31} or
quantum dots,\cite{LeijnsePRB'11, Kondo-Majorana} where the transmission between normal lead and nanowire can be controlled by the intradot Coulomb energy rather than the tunneling barrier.

\vspace{2mm}

\subsection*{Acknowledgments}

This work is supported by Microsoft Q, LPS-CMTC,  and JQI-NSF-PFC. We thank L. Kouwenhoven and C. M. Marcus for discussions.


\begin{thebibliography}{50}%
\makeatletter
\providecommand \@ifxundefined [1]{%
 \@ifx{#1\undefined}
}%
\providecommand \@ifnum [1]{%
 \ifnum #1\expandafter \@firstoftwo
 \else \expandafter \@secondoftwo
 \fi
}%
\providecommand \@ifx [1]{%
 \ifx #1\expandafter \@firstoftwo
 \else \expandafter \@secondoftwo
 \fi
}%
\providecommand \natexlab [1]{#1}%
\providecommand \enquote  [1]{``#1''}%
\providecommand \bibnamefont  [1]{#1}%
\providecommand \bibfnamefont [1]{#1}%
\providecommand \citenamefont [1]{#1}%
\providecommand \href@noop [0]{\@secondoftwo}%
\providecommand \href [0]{\begingroup \@sanitize@url \@href}%
\providecommand \@href[1]{\@@startlink{#1}\@@href}%
\providecommand \@@href[1]{\endgroup#1\@@endlink}%
\providecommand \@sanitize@url [0]{\catcode `\\12\catcode `\$12\catcode
  `\&12\catcode `\#12\catcode `\^12\catcode `\_12\catcode `\%12\relax}%
\providecommand \@@startlink[1]{}%
\providecommand \@@endlink[0]{}%
\providecommand \url  [0]{\begingroup\@sanitize@url \@url }%
\providecommand \@url [1]{\endgroup\@href {#1}{\urlprefix }}%
\providecommand \urlprefix  [0]{URL }%
\providecommand \Eprint [0]{\href }%
\providecommand \doibase [0]{http://dx.doi.org/}%
\providecommand \selectlanguage [0]{\@gobble}%
\providecommand \bibinfo  [0]{\@secondoftwo}%
\providecommand \bibfield  [0]{\@secondoftwo}%
\providecommand \translation [1]{[#1]}%
\providecommand \BibitemOpen [0]{}%
\providecommand \bibitemStop [0]{}%
\providecommand \bibitemNoStop [0]{.\EOS\space}%
\providecommand \EOS [0]{\spacefactor3000\relax}%
\providecommand \BibitemShut  [1]{\csname bibitem#1\endcsname}%
\let\auto@bib@innerbib\@empty
\bibitem [{\citenamefont {Reich}(2012)}]{Reich}%
  \BibitemOpen
  \bibfield  {author} {\bibinfo {author} {\bibfnamefont {E.~S.}\ \bibnamefont
  {Reich}},\ }\href {\doibase 10.1038/483132a} {\bibfield  {journal} {\bibinfo
  {journal} {\nat}\ }\textbf {\bibinfo {volume} {483}},\ \bibinfo {pages} {132}
  (\bibinfo {year} {2012})}\BibitemShut {NoStop}%
\bibitem [{\citenamefont {{Brouwer}}(2012)}]{Brouwer_Science}%
  \BibitemOpen
  \bibfield  {author} {\bibinfo {author} {\bibfnamefont {P.~W.}\ \bibnamefont
  {{Brouwer}}},\ }\href {\doibase 10.1126/science.1223302} {\bibfield
  {journal} {\bibinfo  {journal} {Science}\ }\textbf {\bibinfo {volume}
  {336}},\ \bibinfo {pages} {989} (\bibinfo {year} {2012})}\BibitemShut
  {NoStop}%
\bibitem [{\citenamefont {Wilczek}(2012)}]{Wilczek2012}%
  \BibitemOpen
  \bibfield  {author} {\bibinfo {author} {\bibfnamefont {F.}~\bibnamefont
  {Wilczek}},\ }\href {\doibase 10.1038/486195a} {\bibfield  {journal}
  {\bibinfo  {journal} {\nat}\ }\textbf {\bibinfo {volume} {486}},\ \bibinfo
  {pages} {195} (\bibinfo {year} {2012})}\BibitemShut {NoStop}%
\bibitem [{\citenamefont {Sau}\ \emph {et~al.}(2010{\natexlab{a}})\citenamefont
  {Sau}, \citenamefont {Lutchyn}, \citenamefont {Tewari},\ and\ \citenamefont
  {Das~Sarma}}]{Sau2010}%
  \BibitemOpen
  \bibfield  {author} {\bibinfo {author} {\bibfnamefont {J.~D.}\ \bibnamefont
  {Sau}}, \bibinfo {author} {\bibfnamefont {R.~M.}\ \bibnamefont {Lutchyn}},
  \bibinfo {author} {\bibfnamefont {S.}~\bibnamefont {Tewari}}, \ and\ \bibinfo
  {author} {\bibfnamefont {S.}~\bibnamefont {Das~Sarma}},\ }\href {\doibase
  10.1103/PhysRevLett.104.040502} {\bibfield  {journal} {\bibinfo  {journal}
  {Phys. Rev. Lett.}\ }\textbf {\bibinfo {volume} {104}},\ \bibinfo {pages}
  {040502} (\bibinfo {year} {2010}{\natexlab{a}})}\BibitemShut {NoStop}%
\bibitem [{\citenamefont {Alicea}(2010)}]{Alicea2010}%
  \BibitemOpen
  \bibfield  {author} {\bibinfo {author} {\bibfnamefont {J.}~\bibnamefont
  {Alicea}},\ }\href {\doibase 10.1103/PhysRevB.81.125318} {\bibfield
  {journal} {\bibinfo  {journal} {Phys. Rev. B}\ }\textbf {\bibinfo {volume}
  {81}},\ \bibinfo {pages} {125318} (\bibinfo {year} {2010})}\BibitemShut
  {NoStop}%
\bibitem [{\citenamefont {Lutchyn}\ \emph {et~al.}(2010)\citenamefont
  {Lutchyn}, \citenamefont {Sau},\ and\ \citenamefont
  {Das~Sarma}}]{Lutchyn2010}%
  \BibitemOpen
  \bibfield  {author} {\bibinfo {author} {\bibfnamefont {R.~M.}\ \bibnamefont
  {Lutchyn}}, \bibinfo {author} {\bibfnamefont {J.~D.}\ \bibnamefont {Sau}}, \
  and\ \bibinfo {author} {\bibfnamefont {S.}~\bibnamefont {Das~Sarma}},\ }\href
  {\doibase 10.1103/PhysRevLett.105.077001} {\bibfield  {journal} {\bibinfo
  {journal} {Phys. Rev. Lett.}\ }\textbf {\bibinfo {volume} {105}},\ \bibinfo
  {pages} {077001} (\bibinfo {year} {2010})}\BibitemShut {NoStop}%
\bibitem [{\citenamefont {Oreg}\ \emph {et~al.}(2010)\citenamefont {Oreg},
  \citenamefont {Refael},\ and\ \citenamefont {von Oppen}}]{Oreg2010}%
  \BibitemOpen
  \bibfield  {author} {\bibinfo {author} {\bibfnamefont {Y.}~\bibnamefont
  {Oreg}}, \bibinfo {author} {\bibfnamefont {G.}~\bibnamefont {Refael}}, \ and\
  \bibinfo {author} {\bibfnamefont {F.}~\bibnamefont {von Oppen}},\ }\href
  {\doibase 10.1103/PhysRevLett.105.177002} {\bibfield  {journal} {\bibinfo
  {journal} {Phys. Rev. Lett.}\ }\textbf {\bibinfo {volume} {105}},\ \bibinfo
  {pages} {177002} (\bibinfo {year} {2010})}\BibitemShut {NoStop}%
\bibitem [{\citenamefont {Read}\ and\ \citenamefont {Green}(2000)}]{ReadGreen}%
  \BibitemOpen
  \bibfield  {author} {\bibinfo {author} {\bibfnamefont {N.}~\bibnamefont
  {Read}}\ and\ \bibinfo {author} {\bibfnamefont {D.}~\bibnamefont {Green}},\
  }\href {\doibase 10.1103/PhysRevB.61.10267} {\bibfield  {journal} {\bibinfo
  {journal} {Phys.\ Rev.\ B}\ }\textbf {\bibinfo {volume} {61}},\ \bibinfo
  {pages} {10267} (\bibinfo {year} {2000})}\BibitemShut {NoStop}%
\bibitem [{\citenamefont {Kitaev}(2001)}]{Kitaev:2001}%
  \BibitemOpen
  \bibfield  {author} {\bibinfo {author} {\bibfnamefont {A.~Y.}\ \bibnamefont
  {Kitaev}},\ }\href@noop {} {\bibfield  {journal} {\bibinfo  {journal}
  {Physics-Uspekhi}\ }\textbf {\bibinfo {volume} {44}},\ \bibinfo {pages} {131}
  (\bibinfo {year} {2001})}\BibitemShut {NoStop}%
\bibitem [{\citenamefont {{Das Sarma}}\ \emph {et~al.}(2006)\citenamefont {{Das
  Sarma}}, \citenamefont {Nayak},\ and\ \citenamefont {Tewari}}]{SrRu}%
  \BibitemOpen
  \bibfield  {author} {\bibinfo {author} {\bibfnamefont {S.}~\bibnamefont {{Das
  Sarma}}}, \bibinfo {author} {\bibfnamefont {C.}~\bibnamefont {Nayak}}, \ and\
  \bibinfo {author} {\bibfnamefont {S.}~\bibnamefont {Tewari}},\ }\href@noop {}
  {\bibfield  {journal} {\bibinfo  {journal} {Phys.\ Rev.\ B}\ }\textbf
  {\bibinfo {volume} {73}},\ \bibinfo {pages} {220502(R)} (\bibinfo {year}
  {2006})}\BibitemShut {NoStop}%
\bibitem [{\citenamefont {Fu}\ and\ \citenamefont {Kane}(2008)}]{FuKane}%
  \BibitemOpen
  \bibfield  {author} {\bibinfo {author} {\bibfnamefont {L.}~\bibnamefont
  {Fu}}\ and\ \bibinfo {author} {\bibfnamefont {C.~L.}\ \bibnamefont {Kane}},\
  }\href@noop {} {\bibfield  {journal} {\bibinfo  {journal} {Phys.\ Rev.\
  Lett.}\ }\textbf {\bibinfo {volume} {100}},\ \bibinfo {pages} {096407}
  (\bibinfo {year} {2008})}\BibitemShut {NoStop}%
\bibitem [{\citenamefont {{Moore}}\ and\ \citenamefont
  {{Read}}(1991)}]{Moore1991}%
  \BibitemOpen
  \bibfield  {author} {\bibinfo {author} {\bibfnamefont {G.}~\bibnamefont
  {{Moore}}}\ and\ \bibinfo {author} {\bibfnamefont {N.}~\bibnamefont
  {{Read}}},\ }\href {\doibase 10.1016/0550-3213(91)90407-O} {\bibfield
  {journal} {\bibinfo  {journal} {Nuclear Physics B}\ }\textbf {\bibinfo
  {volume} {360}},\ \bibinfo {pages} {362} (\bibinfo {year}
  {1991})}\BibitemShut {NoStop}%
\bibitem [{\citenamefont {{Nayak}}\ and\ \citenamefont
  {{Wilczek}}(1996)}]{Nayak1996}%
  \BibitemOpen
  \bibfield  {author} {\bibinfo {author} {\bibfnamefont {C.}~\bibnamefont
  {{Nayak}}}\ and\ \bibinfo {author} {\bibfnamefont {F.}~\bibnamefont
  {{Wilczek}}},\ }\href {\doibase 10.1016/0550-3213(96)00430-0} {\bibfield
  {journal} {\bibinfo  {journal} {Nuclear Physics B}\ }\textbf {\bibinfo
  {volume} {479}},\ \bibinfo {pages} {529} (\bibinfo {year}
  {1996})}\BibitemShut {NoStop}%
\bibitem [{\citenamefont {Ivanov}(2001)}]{Ivanov}%
  \BibitemOpen
  \bibfield  {author} {\bibinfo {author} {\bibfnamefont {D.~A.}\ \bibnamefont
  {Ivanov}},\ }\href {\doibase 10.1103/PhysRevLett.86.268} {\bibfield
  {journal} {\bibinfo  {journal} {Phys.\ Rev.\ Lett.}\ }\textbf {\bibinfo
  {volume} {86}},\ \bibinfo {pages} {268} (\bibinfo {year} {2001})}\BibitemShut
  {NoStop}%
\bibitem [{\citenamefont {Nayak}\ \emph {et~al.}(2008)\citenamefont {Nayak},
  \citenamefont {Simon}, \citenamefont {Stern}, \citenamefont {Freedman},\ and\
  \citenamefont {{Das Sarma}}}]{TQCreview}%
  \BibitemOpen
  \bibfield  {author} {\bibinfo {author} {\bibfnamefont {C.}~\bibnamefont
  {Nayak}}, \bibinfo {author} {\bibfnamefont {S.~H.}\ \bibnamefont {Simon}},
  \bibinfo {author} {\bibfnamefont {A.}~\bibnamefont {Stern}}, \bibinfo
  {author} {\bibfnamefont {M.}~\bibnamefont {Freedman}}, \ and\ \bibinfo
  {author} {\bibfnamefont {S.}~\bibnamefont {{Das Sarma}}},\ }\href {\doibase
  10.1103/RevModPhys.80.1083} {\bibfield  {journal} {\bibinfo  {journal} {Rev.\
  Mod.\ Phys.}\ }\textbf {\bibinfo {volume} {80}},\ \bibinfo {pages} {1083}
  (\bibinfo {year} {2008})}\BibitemShut {NoStop}%
\bibitem [{\citenamefont {{Mourik}}\ \emph {et~al.}(2012)\citenamefont
  {{Mourik}}, \citenamefont {{Zuo}}, \citenamefont {{Frolov}}, \citenamefont
  {{Plissard}}, \citenamefont {{Bakkers}},\ and\ \citenamefont
  {{Kouwenhoven}}}]{Mourik2012}%
  \BibitemOpen
  \bibfield  {author} {\bibinfo {author} {\bibfnamefont {V.}~\bibnamefont
  {{Mourik}}}, \bibinfo {author} {\bibfnamefont {K.}~\bibnamefont {{Zuo}}},
  \bibinfo {author} {\bibfnamefont {S.~M.}\ \bibnamefont {{Frolov}}}, \bibinfo
  {author} {\bibfnamefont {S.~R.}\ \bibnamefont {{Plissard}}}, \bibinfo
  {author} {\bibfnamefont {E.~P.~A.~M.}\ \bibnamefont {{Bakkers}}}, \ and\
  \bibinfo {author} {\bibfnamefont {L.~P.}\ \bibnamefont {{Kouwenhoven}}},\
  }\href {\doibase 10.1126/science.1222360} {\bibfield  {journal} {\bibinfo
  {journal} {Science}\ }\textbf {\bibinfo {volume} {336}},\ \bibinfo {pages}
  {1003} (\bibinfo {year} {2012})}\BibitemShut {NoStop}%
\bibitem [{\citenamefont {{Das}}\ \emph {et~al.}(2012)\citenamefont {{Das}},
  \citenamefont {{Ronen}}, \citenamefont {{Most}}, \citenamefont {{Oreg}},
  \citenamefont {{Heiblum}},\ and\ \citenamefont {{Shtrikman}}}]{Das2012}%
  \BibitemOpen
  \bibfield  {author} {\bibinfo {author} {\bibfnamefont {A.}~\bibnamefont
  {{Das}}}, \bibinfo {author} {\bibfnamefont {Y.}~\bibnamefont {{Ronen}}},
  \bibinfo {author} {\bibfnamefont {Y.}~\bibnamefont {{Most}}}, \bibinfo
  {author} {\bibfnamefont {Y.}~\bibnamefont {{Oreg}}}, \bibinfo {author}
  {\bibfnamefont {M.}~\bibnamefont {{Heiblum}}}, \ and\ \bibinfo {author}
  {\bibfnamefont {H.}~\bibnamefont {{Shtrikman}}},\ }\href {\doibase
  10.1038/nphys2479} {\bibfield  {journal} {\bibinfo  {journal} {Nature Phys.}\
  }\textbf {\bibinfo {volume} {8}},\ \bibinfo {pages} {887} (\bibinfo {year}
  {2012})}\BibitemShut {NoStop}%
\bibitem [{\citenamefont {{Deng}}\ \emph {et~al.}(2012)\citenamefont {{Deng}},
  \citenamefont {{Yu}}, \citenamefont {{Huang}}, \citenamefont {{Larsson}},
  \citenamefont {{Caroff}},\ and\ \citenamefont {{Xu}}}]{Deng2012}%
  \BibitemOpen
  \bibfield  {author} {\bibinfo {author} {\bibfnamefont {M.~T.}\ \bibnamefont
  {{Deng}}}, \bibinfo {author} {\bibfnamefont {C.~L.}\ \bibnamefont {{Yu}}},
  \bibinfo {author} {\bibfnamefont {G.~Y.}\ \bibnamefont {{Huang}}}, \bibinfo
  {author} {\bibfnamefont {M.}~\bibnamefont {{Larsson}}}, \bibinfo {author}
  {\bibfnamefont {P.}~\bibnamefont {{Caroff}}}, \ and\ \bibinfo {author}
  {\bibfnamefont {H.~Q.}\ \bibnamefont {{Xu}}},\ }\href {\doibase
  10.1021/nl303758w} {\bibfield  {journal} {\bibinfo  {journal} {Nano Lett.}\
  }\textbf {\bibinfo {volume} {12}},\ \bibinfo {pages} {6414} (\bibinfo {year}
  {2012})}\BibitemShut {NoStop}%
\bibitem [{\citenamefont {{Finck}}\ \emph {et~al.}(2013)\citenamefont
  {{Finck}}, \citenamefont {{Van Harlingen}}, \citenamefont {{Mohseni}},
  \citenamefont {{Jung}},\ and\ \citenamefont {{Li}}}]{Fink2012}%
  \BibitemOpen
  \bibfield  {author} {\bibinfo {author} {\bibfnamefont {A.~D.~K.}\
  \bibnamefont {{Finck}}}, \bibinfo {author} {\bibfnamefont {D.~J.}\
  \bibnamefont {{Van Harlingen}}}, \bibinfo {author} {\bibfnamefont {P.~K.}\
  \bibnamefont {{Mohseni}}}, \bibinfo {author} {\bibfnamefont {K.}~\bibnamefont
  {{Jung}}}, \ and\ \bibinfo {author} {\bibfnamefont {X.}~\bibnamefont
  {{Li}}},\ }\href {\doibase 10.1103/PhysRevLett.110.126406} {\bibfield
  {journal} {\bibinfo  {journal} {\prl}\ }\textbf {\bibinfo {volume} {110}},\
  \bibinfo {pages} {126406} (\bibinfo {year} {2013})}\BibitemShut {NoStop}%
\bibitem [{\citenamefont {Churchill}\ \emph {et~al.}(2013)\citenamefont
  {Churchill}, \citenamefont {Fatemi}, \citenamefont {Grove-Rasmussen},
  \citenamefont {Deng}, \citenamefont {Caroff}, \citenamefont {Xu},\ and\
  \citenamefont {Marcus}}]{Churchill2013}%
  \BibitemOpen
  \bibfield  {author} {\bibinfo {author} {\bibfnamefont {H.~O.~H.}\
  \bibnamefont {Churchill}}, \bibinfo {author} {\bibfnamefont {V.}~\bibnamefont
  {Fatemi}}, \bibinfo {author} {\bibfnamefont {K.}~\bibnamefont
  {Grove-Rasmussen}}, \bibinfo {author} {\bibfnamefont {M.~T.}\ \bibnamefont
  {Deng}}, \bibinfo {author} {\bibfnamefont {P.}~\bibnamefont {Caroff}},
  \bibinfo {author} {\bibfnamefont {H.~Q.}\ \bibnamefont {Xu}}, \ and\ \bibinfo
  {author} {\bibfnamefont {C.~M.}\ \bibnamefont {Marcus}},\ }\href {\doibase
  10.1103/PhysRevB.87.241401} {\bibfield  {journal} {\bibinfo  {journal}
  {\prb}\ }\textbf {\bibinfo {volume} {87}},\ \bibinfo {pages} {241401}
  (\bibinfo {year} {2013})}\BibitemShut {NoStop}%
\bibitem [{\citenamefont {Rokhinson}\ \emph {et~al.}()\citenamefont
  {Rokhinson}, \citenamefont {Liu},\ and\ \citenamefont
  {Furdyna}}]{Rokhinson2012}%
  \BibitemOpen
  \bibfield  {author} {\bibinfo {author} {\bibfnamefont {L.~P.}\ \bibnamefont
  {Rokhinson}}, \bibinfo {author} {\bibfnamefont {X.}~\bibnamefont {Liu}}, \
  and\ \bibinfo {author} {\bibfnamefont {J.~K.}\ \bibnamefont {Furdyna}},\
  }\href@noop {} {}\bibinfo {howpublished} {{\it Nature Physics} {\bf 8}, 795
  (2012)}\BibitemShut {NoStop}%
\bibitem [{\citenamefont {Sengupta}\ \emph {et~al.}(2001)\citenamefont
  {Sengupta}, \citenamefont {\ifmmode \check{Z}\else
  \v{Z}\fi{}uti\ifmmode~\acute{c}\else \'{c}\fi{}}, \citenamefont {Kwon},
  \citenamefont {Yakovenko},\ and\ \citenamefont
  {Das~Sarma}}]{ZeroBiasAnomaly0}%
  \BibitemOpen
  \bibfield  {author} {\bibinfo {author} {\bibfnamefont {K.}~\bibnamefont
  {Sengupta}}, \bibinfo {author} {\bibfnamefont {I.}~\bibnamefont {\ifmmode
  \check{Z}\else \v{Z}\fi{}uti\ifmmode~\acute{c}\else \'{c}\fi{}}}, \bibinfo
  {author} {\bibfnamefont {H.-J.}\ \bibnamefont {Kwon}}, \bibinfo {author}
  {\bibfnamefont {V.~M.}\ \bibnamefont {Yakovenko}}, \ and\ \bibinfo {author}
  {\bibfnamefont {S.}~\bibnamefont {Das~Sarma}},\ }\href {\doibase
  10.1103/PhysRevB.63.144531} {\bibfield  {journal} {\bibinfo  {journal}
  {\prb}\ }\textbf {\bibinfo {volume} {63}},\ \bibinfo {pages} {144531}
  (\bibinfo {year} {2001})}\BibitemShut {NoStop}%
\bibitem [{\citenamefont {Bolech}\ and\ \citenamefont
  {Demler}(2007)}]{ZeroBiasAnomaly1}%
  \BibitemOpen
  \bibfield  {author} {\bibinfo {author} {\bibfnamefont {C.~J.}\ \bibnamefont
  {Bolech}}\ and\ \bibinfo {author} {\bibfnamefont {E.}~\bibnamefont
  {Demler}},\ }\href {\doibase 10.1103/PhysRevLett.98.237002} {\bibfield
  {journal} {\bibinfo  {journal} {\prl}\ }\textbf {\bibinfo {volume} {98}},\
  \bibinfo {pages} {237002} (\bibinfo {year} {2007})}\BibitemShut {NoStop}%
\bibitem [{\citenamefont {Nilsson}\ \emph {et~al.}(2008)\citenamefont
  {Nilsson}, \citenamefont {Akhmerov},\ and\ \citenamefont
  {Beenakker}}]{ZeroBiasAnomaly2}%
  \BibitemOpen
  \bibfield  {author} {\bibinfo {author} {\bibfnamefont {J.}~\bibnamefont
  {Nilsson}}, \bibinfo {author} {\bibfnamefont {A.~R.}\ \bibnamefont
  {Akhmerov}}, \ and\ \bibinfo {author} {\bibfnamefont {C.~W.~J.}\ \bibnamefont
  {Beenakker}},\ }\href {\doibase 10.1103/PhysRevLett.101.120403} {\bibfield
  {journal} {\bibinfo  {journal} {\prl}\ }\textbf {\bibinfo {volume} {101}},\
  \bibinfo {pages} {120403} (\bibinfo {year} {2008})}\BibitemShut {NoStop}%
\bibitem [{\citenamefont {Law}\ \emph {et~al.}(2009)\citenamefont {Law},
  \citenamefont {Lee},\ and\ \citenamefont {Ng}}]{ZeroBiasAnomaly3}%
  \BibitemOpen
  \bibfield  {author} {\bibinfo {author} {\bibfnamefont {K.~T.}\ \bibnamefont
  {Law}}, \bibinfo {author} {\bibfnamefont {P.~A.}\ \bibnamefont {Lee}}, \ and\
  \bibinfo {author} {\bibfnamefont {T.~K.}\ \bibnamefont {Ng}},\ }\href
  {\doibase 10.1103/PhysRevLett.103.237001} {\bibfield  {journal} {\bibinfo
  {journal} {Phys.\ Rev.\ Lett.}\ }\textbf {\bibinfo {volume} {103}},\ \bibinfo
  {pages} {237001} (\bibinfo {year} {2009})}\BibitemShut {NoStop}%
\bibitem [{\citenamefont {Sau}\ \emph {et~al.}(2010{\natexlab{b}})\citenamefont
  {Sau}, \citenamefont {Tewari}, \citenamefont {Lutchyn}, \citenamefont
  {Stanescu},\ and\ \citenamefont {Das~Sarma}}]{ZeroBiasAnomaly31}%
  \BibitemOpen
  \bibfield  {author} {\bibinfo {author} {\bibfnamefont {J.~D.}\ \bibnamefont
  {Sau}}, \bibinfo {author} {\bibfnamefont {S.}~\bibnamefont {Tewari}},
  \bibinfo {author} {\bibfnamefont {R.~M.}\ \bibnamefont {Lutchyn}}, \bibinfo
  {author} {\bibfnamefont {T.~D.}\ \bibnamefont {Stanescu}}, \ and\ \bibinfo
  {author} {\bibfnamefont {S.}~\bibnamefont {Das~Sarma}},\ }\href {\doibase
  10.1103/PhysRevB.82.214509} {\bibfield  {journal} {\bibinfo  {journal}
  {\prb}\ }\textbf {\bibinfo {volume} {82}},\ \bibinfo {pages} {214509}
  (\bibinfo {year} {2010}{\natexlab{b}})}\BibitemShut {NoStop}%
\bibitem [{\citenamefont {Flensberg}(2010)}]{ZeroBiasAnomaly4}%
  \BibitemOpen
  \bibfield  {author} {\bibinfo {author} {\bibfnamefont {K.}~\bibnamefont
  {Flensberg}},\ }\href {\doibase 10.1103/PhysRevB.82.180516} {\bibfield
  {journal} {\bibinfo  {journal} {\prb}\ }\textbf {\bibinfo {volume} {82}},\
  \bibinfo {pages} {180516} (\bibinfo {year} {2010})}\BibitemShut {NoStop}%
\bibitem [{\citenamefont {Golub}\ and\ \citenamefont
  {Horovitz}(2011)}]{ZeroBiasAnomaly5}%
  \BibitemOpen
  \bibfield  {author} {\bibinfo {author} {\bibfnamefont {A.}~\bibnamefont
  {Golub}}\ and\ \bibinfo {author} {\bibfnamefont {B.}~\bibnamefont
  {Horovitz}},\ }\href {\doibase 10.1103/PhysRevB.83.153415} {\bibfield
  {journal} {\bibinfo  {journal} {\prb}\ }\textbf {\bibinfo {volume} {83}},\
  \bibinfo {pages} {153415} (\bibinfo {year} {2011})}\BibitemShut {NoStop}%
\bibitem [{\citenamefont {Wimmer}\ \emph {et~al.}(2011)\citenamefont {Wimmer},
  \citenamefont {Akhmerov}, \citenamefont {Dahlhaus},\ and\ \citenamefont
  {Beenakker}}]{ZeroBiasAnomaly6}%
  \BibitemOpen
  \bibfield  {author} {\bibinfo {author} {\bibfnamefont {M.}~\bibnamefont
  {Wimmer}}, \bibinfo {author} {\bibfnamefont {A.~R.}\ \bibnamefont
  {Akhmerov}}, \bibinfo {author} {\bibfnamefont {J.~P.}\ \bibnamefont
  {Dahlhaus}}, \ and\ \bibinfo {author} {\bibfnamefont {C.~W.~J.}\ \bibnamefont
  {Beenakker}},\ }\href {\doibase 10.1088/1367-2630/13/5/053016} {\bibfield
  {journal} {\bibinfo  {journal} {New J. Phys.}\ }\textbf {\bibinfo {volume}
  {13}},\ \bibinfo {pages} {053016} (\bibinfo {year} {2011})}\BibitemShut
  {NoStop}%
\bibitem [{\citenamefont {{Lutchyn}}\ \emph {et~al.}(2011)\citenamefont
  {{Lutchyn}}, \citenamefont {{Stanescu}},\ and\ \citenamefont {{Das
  Sarma}}}]{1DwiresLutchyn2}%
  \BibitemOpen
  \bibfield  {author} {\bibinfo {author} {\bibfnamefont {R.~M.}\ \bibnamefont
  {{Lutchyn}}}, \bibinfo {author} {\bibfnamefont {T.~D.}\ \bibnamefont
  {{Stanescu}}}, \ and\ \bibinfo {author} {\bibfnamefont {S.}~\bibnamefont
  {{Das Sarma}}},\ }\href {\doibase 10.1103/PhysRevLett.106.127001} {\bibfield
  {journal} {\bibinfo  {journal} {\prl}\ }\textbf {\bibinfo {volume} {106}},\
  \bibinfo {eid} {127001} (\bibinfo {year} {2011})}\BibitemShut {NoStop}%
\bibitem [{\citenamefont {Stanescu}\ \emph
  {et~al.}(2011{\natexlab{a}})\citenamefont {Stanescu}, \citenamefont
  {Lutchyn},\ and\ \citenamefont {Das~Sarma}}]{ZeroBiasAnomaly61}%
  \BibitemOpen
  \bibfield  {author} {\bibinfo {author} {\bibfnamefont {T.~D.}\ \bibnamefont
  {Stanescu}}, \bibinfo {author} {\bibfnamefont {R.~M.}\ \bibnamefont
  {Lutchyn}}, \ and\ \bibinfo {author} {\bibfnamefont {S.}~\bibnamefont
  {Das~Sarma}},\ }\href {\doibase 10.1103/PhysRevB.84.144522} {\bibfield
  {journal} {\bibinfo  {journal} {\prb}\ }\textbf {\bibinfo {volume} {84}},\
  \bibinfo {pages} {144522} (\bibinfo {year} {2011}{\natexlab{a}})}\BibitemShut
  {NoStop}%
\bibitem [{\citenamefont {Qu}\ \emph {et~al.}(2011)\citenamefont {Qu},
  \citenamefont {Zhang}, \citenamefont {Mao},\ and\ \citenamefont
  {Zhang}}]{ZeroBiasAnomaly7}%
  \BibitemOpen
  \bibfield  {author} {\bibinfo {author} {\bibfnamefont {C.}~\bibnamefont
  {Qu}}, \bibinfo {author} {\bibfnamefont {Y.}~\bibnamefont {Zhang}}, \bibinfo
  {author} {\bibfnamefont {L.}~\bibnamefont {Mao}}, \ and\ \bibinfo {author}
  {\bibfnamefont {C.}~\bibnamefont {Zhang}},\ }\href@noop {} {\bibfield
  {journal} {\bibinfo  {journal} {arXiv:1109.4108}\ } (\bibinfo {year}
  {2011})}\BibitemShut {NoStop}%
\bibitem [{\citenamefont {Alicea}\ \emph {et~al.}(2011)\citenamefont {Alicea},
  \citenamefont {Oreg}, \citenamefont {Refael}, \citenamefont {von Oppen},\
  and\ \citenamefont {Fisher}}]{AliceaBraiding}%
  \BibitemOpen
  \bibfield  {author} {\bibinfo {author} {\bibfnamefont {J.}~\bibnamefont
  {Alicea}}, \bibinfo {author} {\bibfnamefont {Y.}~\bibnamefont {Oreg}},
  \bibinfo {author} {\bibfnamefont {G.}~\bibnamefont {Refael}}, \bibinfo
  {author} {\bibfnamefont {F.}~\bibnamefont {von Oppen}}, \ and\ \bibinfo
  {author} {\bibfnamefont {M.~P.~A.}\ \bibnamefont {Fisher}},\ }\href {\doibase
  10.1038/nphys1915} {\bibfield  {journal} {\bibinfo  {journal} {Nature Phys.}\
  }\textbf {\bibinfo {volume} {7}},\ \bibinfo {pages} {412} (\bibinfo {year}
  {2011})}\BibitemShut {NoStop}%
\bibitem [{\citenamefont {Sau}\ \emph {et~al.}(2010{\natexlab{c}})\citenamefont
  {Sau}, \citenamefont {Tewari},\ and\ \citenamefont {{Das
  Sarma}}}]{SauWireNetwork}%
  \BibitemOpen
  \bibfield  {author} {\bibinfo {author} {\bibfnamefont {J.~D.}\ \bibnamefont
  {Sau}}, \bibinfo {author} {\bibfnamefont {S.}~\bibnamefont {Tewari}}, \ and\
  \bibinfo {author} {\bibfnamefont {S.}~\bibnamefont {{Das Sarma}}},\ }\href
  {\doibase 10.1103/PhysRevA.82.052322} {\bibfield  {journal} {\bibinfo
  {journal} {Phys. Rev. A}\ }\textbf {\bibinfo {volume} {82}},\ \bibinfo
  {pages} {052322} (\bibinfo {year} {2010}{\natexlab{c}})}\BibitemShut
  {NoStop}%
\bibitem [{\citenamefont {Clarke}\ \emph {et~al.}(2011)\citenamefont {Clarke},
  \citenamefont {Sau},\ and\ \citenamefont {Tewari}}]{ClarkeBraiding}%
  \BibitemOpen
  \bibfield  {author} {\bibinfo {author} {\bibfnamefont {D.~J.}\ \bibnamefont
  {Clarke}}, \bibinfo {author} {\bibfnamefont {J.~D.}\ \bibnamefont {Sau}}, \
  and\ \bibinfo {author} {\bibfnamefont {S.}~\bibnamefont {Tewari}},\ }\href
  {\doibase 10.1103/PhysRevB.84.035120} {\bibfield  {journal} {\bibinfo
  {journal} {\prb}\ }\textbf {\bibinfo {volume} {84}},\ \bibinfo {pages}
  {035120} (\bibinfo {year} {2011})}\BibitemShut {NoStop}%
\bibitem [{\citenamefont {Bonderson}\ and\ \citenamefont
  {Lutchyn}(2011)}]{TopologicalQuantumBus}%
  \BibitemOpen
  \bibfield  {author} {\bibinfo {author} {\bibfnamefont {P.}~\bibnamefont
  {Bonderson}}\ and\ \bibinfo {author} {\bibfnamefont {R.~M.}\ \bibnamefont
  {Lutchyn}},\ }\href {\doibase 10.1103/PhysRevLett.106.130505} {\bibfield
  {journal} {\bibinfo  {journal} {\prl}\ }\textbf {\bibinfo {volume} {106}},\
  \bibinfo {pages} {130505} (\bibinfo {year} {2011})}\BibitemShut {NoStop}%
\bibitem [{\citenamefont {van Heck}\ \emph {et~al.}(2012)\citenamefont {van
  Heck}, \citenamefont {Akhmerov}, \citenamefont {Hassler}, \citenamefont
  {Burrello},\ and\ \citenamefont {Beenakker}}]{BraidingWithoutTransport}%
  \BibitemOpen
  \bibfield  {author} {\bibinfo {author} {\bibfnamefont {B.}~\bibnamefont {van
  Heck}}, \bibinfo {author} {\bibfnamefont {A.~R.}\ \bibnamefont {Akhmerov}},
  \bibinfo {author} {\bibfnamefont {F.}~\bibnamefont {Hassler}}, \bibinfo
  {author} {\bibfnamefont {M.}~\bibnamefont {Burrello}}, \ and\ \bibinfo
  {author} {\bibfnamefont {C.~W.~J.}\ \bibnamefont {Beenakker}},\ }\href
  {\doibase 10.1088/1367-2630/14/3/035019} {\bibfield  {journal} {\bibinfo
  {journal} {New J. Phys.}\ }\textbf {\bibinfo {volume} {14}},\ \bibinfo
  {pages} {035019} (\bibinfo {year} {2012})}\BibitemShut {NoStop}%
\bibitem [{\citenamefont {{Cheng}}\ \emph {et~al.}(2012)\citenamefont
  {{Cheng}}, \citenamefont {{Lutchyn}},\ and\ \citenamefont {{Das
  Sarma}}}]{ChengPRB'12}%
  \BibitemOpen
  \bibfield  {author} {\bibinfo {author} {\bibfnamefont {M.}~\bibnamefont
  {{Cheng}}}, \bibinfo {author} {\bibfnamefont {R.~M.}\ \bibnamefont
  {{Lutchyn}}}, \ and\ \bibinfo {author} {\bibfnamefont {S.}~\bibnamefont {{Das
  Sarma}}},\ }\href {\doibase 10.1103/PhysRevB.85.165124} {\bibfield  {journal}
  {\bibinfo  {journal} {\prb}\ }\textbf {\bibinfo {volume} {85}},\ \bibinfo
  {eid} {165124} (\bibinfo {year} {2012})},\ \Eprint
  {http://arxiv.org/abs/1112.3662} {arXiv:1112.3662 [cond-mat.supr-con]}
  \BibitemShut {NoStop}%
\bibitem [{\citenamefont {{Akhmerov}}\ \emph {et~al.}(2011)\citenamefont
  {{Akhmerov}}, \citenamefont {{Dahlhaus}}, \citenamefont {{Hassler}},
  \citenamefont {{Wimmer}},\ and\ \citenamefont
  {{Beenakker}}}]{AkhmerovPRL'11}%
  \BibitemOpen
  \bibfield  {author} {\bibinfo {author} {\bibfnamefont {A.~R.}\ \bibnamefont
  {{Akhmerov}}}, \bibinfo {author} {\bibfnamefont {J.~P.}\ \bibnamefont
  {{Dahlhaus}}}, \bibinfo {author} {\bibfnamefont {F.}~\bibnamefont
  {{Hassler}}}, \bibinfo {author} {\bibfnamefont {M.}~\bibnamefont {{Wimmer}}},
  \ and\ \bibinfo {author} {\bibfnamefont {C.~W.~J.}\ \bibnamefont
  {{Beenakker}}},\ }\href {\doibase 10.1103/PhysRevLett.106.057001} {\bibfield
  {journal} {\bibinfo  {journal} {Physical Review Letters}\ }\textbf {\bibinfo
  {volume} {106}},\ \bibinfo {eid} {057001} (\bibinfo {year} {2011})},\ \Eprint
  {http://arxiv.org/abs/1009.5542} {arXiv:1009.5542 [cond-mat.mes-hall]}
  \BibitemShut {NoStop}%
\bibitem [{\citenamefont {{Prada}}\ \emph {et~al.}(2012)\citenamefont
  {{Prada}}, \citenamefont {{San-Jose}},\ and\ \citenamefont
  {{Aguado}}}]{PradaPRB'12}%
  \BibitemOpen
  \bibfield  {author} {\bibinfo {author} {\bibfnamefont {E.}~\bibnamefont
  {{Prada}}}, \bibinfo {author} {\bibfnamefont {P.}~\bibnamefont {{San-Jose}}},
  \ and\ \bibinfo {author} {\bibfnamefont {R.}~\bibnamefont {{Aguado}}},\
  }\href {\doibase 10.1103/PhysRevB.86.180503} {\bibfield  {journal} {\bibinfo
  {journal} {\prb}\ }\textbf {\bibinfo {volume} {86}},\ \bibinfo {eid} {180503}
  (\bibinfo {year} {2012})},\ \Eprint {http://arxiv.org/abs/1203.4488}
  {arXiv:1203.4488 [cond-mat.mes-hall]} \BibitemShut {NoStop}%
\bibitem [{\citenamefont {{Pientka}}\ \emph {et~al.}(2012)\citenamefont
  {{Pientka}}, \citenamefont {{Kells}}, \citenamefont {{Romito}}, \citenamefont
  {{Brouwer}},\ and\ \citenamefont {{von Oppen}}}]{PientkaPRL'12}%
  \BibitemOpen
  \bibfield  {author} {\bibinfo {author} {\bibfnamefont {F.}~\bibnamefont
  {{Pientka}}}, \bibinfo {author} {\bibfnamefont {G.}~\bibnamefont {{Kells}}},
  \bibinfo {author} {\bibfnamefont {A.}~\bibnamefont {{Romito}}}, \bibinfo
  {author} {\bibfnamefont {P.~W.}\ \bibnamefont {{Brouwer}}}, \ and\ \bibinfo
  {author} {\bibfnamefont {F.}~\bibnamefont {{von Oppen}}},\ }\href {\doibase
  10.1103/PhysRevLett.109.227006} {\bibfield  {journal} {\bibinfo  {journal}
  {Physical Review Letters}\ }\textbf {\bibinfo {volume} {109}},\ \bibinfo
  {eid} {227006} (\bibinfo {year} {2012})},\ \Eprint
  {http://arxiv.org/abs/1206.0723} {arXiv:1206.0723 [cond-mat.mes-hall]}
  \BibitemShut {NoStop}%
\bibitem [{\citenamefont {Rainis}\ \emph {et~al.}(2013)\citenamefont {Rainis},
  \citenamefont {Trifunovic}, \citenamefont {Klinovaja},\ and\ \citenamefont
  {Loss}}]{RainisPRB'13}%
  \BibitemOpen
  \bibfield  {author} {\bibinfo {author} {\bibfnamefont {D.}~\bibnamefont
  {Rainis}}, \bibinfo {author} {\bibfnamefont {L.}~\bibnamefont {Trifunovic}},
  \bibinfo {author} {\bibfnamefont {J.}~\bibnamefont {Klinovaja}}, \ and\
  \bibinfo {author} {\bibfnamefont {D.}~\bibnamefont {Loss}},\ }\href {\doibase
  10.1103/PhysRevB.87.024515} {\bibfield  {journal} {\bibinfo  {journal} {Phys.
  Rev. B}\ }\textbf {\bibinfo {volume} {87}},\ \bibinfo {pages} {024515}
  (\bibinfo {year} {2013})}\BibitemShut {NoStop}%
\bibitem [{\citenamefont {Blonder}\ \emph {et~al.}(1982)\citenamefont
  {Blonder}, \citenamefont {Tinkham},\ and\ \citenamefont
  {Klapwijk}}]{BTK}%
  \BibitemOpen
  \bibfield  {author} {\bibinfo {author} {\bibfnamefont {G.~E.}\ \bibnamefont
  {Blonder}}, \bibinfo {author} {\bibfnamefont {M.}~\bibnamefont {Tinkham}}, \
  and\ \bibinfo {author} {\bibfnamefont {T.~M.}\ \bibnamefont {Klapwijk}},\
  }\href {\doibase 10.1103/PhysRevB.25.4515} {\bibfield  {journal} {\bibinfo
  {journal} {Phys. Rev. B}\ }\textbf {\bibinfo {volume} {25}},\ \bibinfo
  {pages} {4515} (\bibinfo {year} {1982})}\BibitemShut {NoStop}%
\bibitem [{\citenamefont {Takei}\ \emph {et~al.}(2013)\citenamefont {Takei},
  \citenamefont {Fregoso}, \citenamefont {Hui}, \citenamefont {Lobos},\ and\
  \citenamefont {Das~Sarma}}]{TakeiPRL'13}%
  \BibitemOpen
  \bibfield  {author} {\bibinfo {author} {\bibfnamefont {S.}~\bibnamefont
  {Takei}}, \bibinfo {author} {\bibfnamefont {B.~M.}\ \bibnamefont {Fregoso}},
  \bibinfo {author} {\bibfnamefont {H.-Y.}\ \bibnamefont {Hui}}, \bibinfo
  {author} {\bibfnamefont {A.~M.}\ \bibnamefont {Lobos}}, \ and\ \bibinfo
  {author} {\bibfnamefont {S.}~\bibnamefont {Das~Sarma}},\ }\href {\doibase
  10.1103/PhysRevLett.110.186803} {\bibfield  {journal} {\bibinfo  {journal}
  {Phys. Rev. Lett.}\ }\textbf {\bibinfo {volume} {110}},\ \bibinfo {pages}
  {186803} (\bibinfo {year} {2013})}\BibitemShut {NoStop}%
\bibitem [{\citenamefont {Stanescu}\ \emph
  {et~al.}(2011{\natexlab{b}})\citenamefont {Stanescu}, \citenamefont
  {Lutchyn},\ and\ \citenamefont {Das~Sarma}}]{Stanescu2011}%
  \BibitemOpen
  \bibfield  {author} {\bibinfo {author} {\bibfnamefont {T.~D.}\ \bibnamefont
  {Stanescu}}, \bibinfo {author} {\bibfnamefont {R.~M.}\ \bibnamefont
  {Lutchyn}}, \ and\ \bibinfo {author} {\bibfnamefont {S.}~\bibnamefont
  {Das~Sarma}},\ }\href {\doibase 10.1103/PhysRevB.84.144522} {\bibfield
  {journal} {\bibinfo  {journal} {Phys. Rev. B}\ }\textbf {\bibinfo {volume}
  {84}},\ \bibinfo {pages} {144522} (\bibinfo {year}
  {2011}{\natexlab{b}})}\BibitemShut {NoStop}%
\bibitem [{\citenamefont {Stanescu}\ and\ \citenamefont
  {Das~Sarma}(2013)}]{Stanescu2013a}%
  \BibitemOpen
  \bibfield  {author} {\bibinfo {author} {\bibfnamefont {T.~D.}\ \bibnamefont
  {Stanescu}}\ and\ \bibinfo {author} {\bibfnamefont {S.}~\bibnamefont
  {Das~Sarma}},\ }\href {\doibase 10.1103/PhysRevB.87.180504} {\bibfield
  {journal} {\bibinfo  {journal} {Phys. Rev. B}\ }\textbf {\bibinfo {volume}
  {87}},\ \bibinfo {pages} {180504} (\bibinfo {year} {2013})}\BibitemShut
  {NoStop}%
\bibitem [{\citenamefont {Stanescu}\ and\ \citenamefont
  {Tewari}()}]{Stanescu2013b}%
  \BibitemOpen
  \bibfield  {author} {\bibinfo {author} {\bibfnamefont {T.~D.}\ \bibnamefont
  {Stanescu}}\ and\ \bibinfo {author} {\bibfnamefont {S.}~\bibnamefont
  {Tewari}},\ }\href@noop {} {}\bibinfo {howpublished} {J. Phys.: Condens.
  Matter {\bf 25}, 233201 (2013)}\BibitemShut {NoStop}%
\bibitem [{\citenamefont {Lin}\ \emph {et~al.}(2012)\citenamefont {Lin},
  \citenamefont {Sau},\ and\ \citenamefont {Das~Sarma}}]{ScattMat2}%
  \BibitemOpen
  \bibfield  {author} {\bibinfo {author} {\bibfnamefont {C.-H.}\ \bibnamefont
  {Lin}}, \bibinfo {author} {\bibfnamefont {J.~D.}\ \bibnamefont {Sau}}, \ and\
  \bibinfo {author} {\bibfnamefont {S.}~\bibnamefont {Das~Sarma}},\ }\href
  {\doibase 10.1103/PhysRevB.86.224511} {\bibfield  {journal} {\bibinfo
  {journal} {Phys. Rev. B}\ }\textbf {\bibinfo {volume} {86}},\ \bibinfo
  {pages} {224511} (\bibinfo {year} {2012})}\BibitemShut {NoStop}%
\bibitem [{\citenamefont {Cheng}\ \emph {et~al.}(2009)\citenamefont {Cheng},
  \citenamefont {Lutchyn}, \citenamefont {Galitski},\ and\ \citenamefont
  {Das~Sarma}}]{Meng_splitting}%
  \BibitemOpen
  \bibfield  {author} {\bibinfo {author} {\bibfnamefont {M.}~\bibnamefont
  {Cheng}}, \bibinfo {author} {\bibfnamefont {R.~M.}\ \bibnamefont {Lutchyn}},
  \bibinfo {author} {\bibfnamefont {V.}~\bibnamefont {Galitski}}, \ and\
  \bibinfo {author} {\bibfnamefont {S.}~\bibnamefont {Das~Sarma}},\ }\href
  {\doibase 10.1103/PhysRevLett.103.107001} {\bibfield  {journal} {\bibinfo
  {journal} {\prl}\ }\textbf {\bibinfo {volume} {103}},\ \bibinfo {pages}
  {107001} (\bibinfo {year} {2009})}\BibitemShut {NoStop}%
\bibitem [{\citenamefont {Das~Sarma}\ \emph {et~al.}(2012)\citenamefont
  {Das~Sarma}, \citenamefont {Sau},\ and\ \citenamefont
  {Stanescu}}]{DSarma2012}%
  \BibitemOpen
  \bibfield  {author} {\bibinfo {author} {\bibfnamefont {S.}~\bibnamefont
  {Das~Sarma}}, \bibinfo {author} {\bibfnamefont {J.~D.}\ \bibnamefont {Sau}},
  \ and\ \bibinfo {author} {\bibfnamefont {T.~D.}\ \bibnamefont {Stanescu}},\
  }\href {\doibase 10.1103/PhysRevB.86.220506} {\bibfield  {journal} {\bibinfo
  {journal} {Phys. Rev. B}\ }\textbf {\bibinfo {volume} {86}},\ \bibinfo
  {pages} {220506} (\bibinfo {year} {2012})}\BibitemShut {NoStop}%
\bibitem [{\citenamefont {{Leijnse}}\ and\ \citenamefont
  {{Flensberg}}(2011)}]{LeijnsePRB'11}%
  \BibitemOpen
  \bibfield  {author} {\bibinfo {author} {\bibfnamefont {M.}~\bibnamefont
  {{Leijnse}}}\ and\ \bibinfo {author} {\bibfnamefont {K.}~\bibnamefont
  {{Flensberg}}},\ }\href {\doibase 10.1103/PhysRevB.84.140501} {\bibfield
  {journal} {\bibinfo  {journal} {\prb}\ }\textbf {\bibinfo {volume} {84}},\
  \bibinfo {eid} {140501} (\bibinfo {year} {2011})}\BibitemShut {NoStop}%
\bibitem [{\citenamefont {{Cheng}}\ \emph {et~al.}(2013)\citenamefont
  {{Cheng}}, \citenamefont {{Becker}}, \citenamefont {{Bauer}},\ and\
  \citenamefont {{Lutchyn}}}]{Kondo-Majorana}%
  \BibitemOpen
  \bibfield  {author} {\bibinfo {author} {\bibfnamefont {M.}~\bibnamefont
  {{Cheng}}}, \bibinfo {author} {\bibfnamefont {M.}~\bibnamefont {{Becker}}},
  \bibinfo {author} {\bibfnamefont {B.}~\bibnamefont {{Bauer}}}, \ and\
  \bibinfo {author} {\bibfnamefont {R.~M.}\ \bibnamefont {{Lutchyn}}},\
  }\href@noop {} {\bibfield  {journal} {\bibinfo  {journal} {ArXiv e-prints}\ }
  (\bibinfo {year} {2013})},\ \Eprint {http://arxiv.org/abs/1308.4156}
  {arXiv:1308.4156 [cond-mat.mes-hall]} \BibitemShut {NoStop}%
\end{thebibliography}


%

\end{document}